\documentclass[journal=jacsat,manuscript=article,layout=twocolumn]{achemso}
\setkeys{acs}{keywords = true}
\usepackage[version=3]{mhchem} 
\usepackage[hidelinks]{hyperref}
\usepackage{xcolor}
\usepackage[normalem]{ulem}

\let\oldmaketitle\maketitle
\let\maketitle\relax

\author{Lilian Witthauer}
\author{Juan Pedro Cascales}
\author{Emmanuel Roussakis}
\author{Xiaolei Li}
\author{Avery Goss}
\author{Yenyu Chen}
\author{Conor L. Evans}
\email{Evans.Conor@mgh.harvard.edu}
\affiliation[Wellman Center]{Wellman Center for Photomedicine, Massachusetts General Hospital, Harvard Medical School, Charlestown, 02129, MA}

\title[]
  {Portable oxygen-sensing device for the improved assessment of compartment syndrome and other hypoxia related conditions}


\begin{document}

\begin{tocentry}
\centering
\includegraphics[width=7cm]{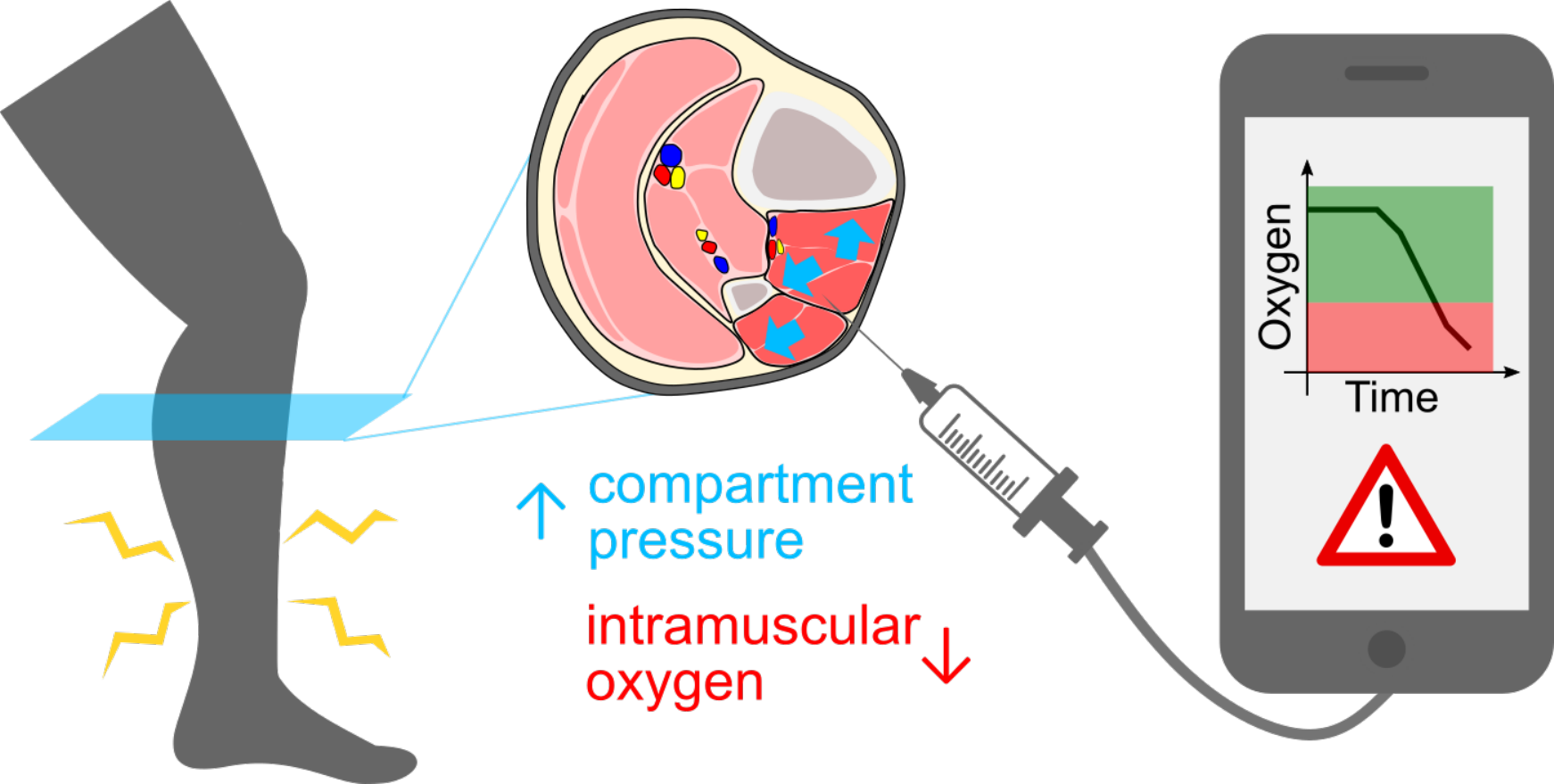}

\end{tocentry}

\twocolumn[
\begin{@twocolumnfalse}
\oldmaketitle
\begin{abstract}
\footnotesize

Measurement of intramuscular oxygen could play a key role in the early diagnosis of acute compartment syndrome, a common condition occurring after severe trauma leading to ischemia and long term consequences including rhabdomyolysis, limb loss, and death. However, to date there is no existing oxygen sensor approved for such a purpose. In order to address the need to improve the assessment of compartment syndrome, a portable fiber-optic device for intramuscular oxygen measurements was developed. The device is based on phosphorescence quenching, where the tip of an optical fiber was coated with a poly(propyl methacrylate) (PPMA) matrix containing a brightly emitting Pt(II)-core phorphyrin. The optoelectronic circuit is highly portable and is based on a micro-spectrometer and a microcontroller read-out with a smartphone. Results from an \textit{in vivo} tourniquet porcine model show that the sensor is sensitive across the physiological oxygen partial pressure range of 0-80~mmHg and exhibits an appropriate and reproducible response to changes in intramuscular oxygen. \textcolor{black}{A commercial laboratory oxygen sensor based on a lifetime measurement did not respond as expected.}
\end{abstract}
\end{@twocolumnfalse}
]
Acute compartment syndrome (ACS) is a condition where muscle ischemia occurs as a consequence of severe injury, which can arise from numerous forms of trauma. \citeauthor{McQueen2000-ot} found that tibia fracture, soft tissue injury, and radial fractures after traffic or sports accidents are the leading cause of ACS in civilian populations \cite{McQueen2000-ot}, although it can appear after nonaccidental causes like bleeding disorder or diabetes mellitus \cite{Raza2015-ey}. ACS also plays an important role in military medicine when it occurs after polytrauma, blunt or crushing injuries from explosions, and tourniquet application \cite{Gordon2018-ll}. The condition mainly affects young men below 35 with an incidence of 7.3 in 100,000 patients versus 0.7 in 100,000 for women.\cite{McQueen2000-ot} 

The pathophysiology of ACS is described as an increase in pressure within the confined space of a compartment \cite{Tollens1998-kb}. The increased pressure results in a decrease in perfusion pressure that impairs blood supply and drainage, resulting in tissue hypoxia, tissue necrosis, and nerve damage. \textcolor{black}{Since tissue necrosis already occurs within 6-12 hours following hypoxemia, ACS should be treated immediately after diagnosis\cite{Von_Keudell2015-lq}. 
Patients with ACS treated after the 6-12 hour time frame were shown to have higher risk of developing worse clinical outcomes such as} loss of function, amputation of the limb or life-threatening conditions \cite{Zhang2020-kn,Gordon2018-cx}, which underscores the importance of early diagnosis of this condition. 

Currently, the clinical standard \textcolor{black}{treatment} for compartment syndrome is a fasciomotomy, where deep incisions are performed to release the pressure, resulting in severe scaring and chronic pain. \cite{Mauffrey2019-sy}.

The current diagnostic standard for ACS is focused on a neurovascular integrity assessment based on five clinical signs: pain, pallor, paresthesia, pulselessness, and paralysis \cite{Gordon2018-ll}. Pain out of proportion was observed to be the most important clinical sign \cite{Von_Keudell2015-lq, Carrasco2015-vz}, however, pain is not specific to ACS and can arise from other injuries. It is also worth noting that the communication of severe pain is not possible in unconscious patients.

In some cases, the neurovascular assessment is complemented with a measurement of compartment pressure (CP) or perfusion pressure ($\Delta$p) \cite{Whitesides1975-jz,Halanski2015-qt,McMillan2019-bu,Guo2019-nc,Mauffrey2019-sy}. CP measurements can be done using a simple arterial line transducer or with proprietary devices; for example, the C2Dx STIC pressure monitor (formerly the Stryker monitor) or the Millar Solid State Pressure Sensor.
In practice, CP measurements are not widely used due to the high costs involved with the proprietary devices and the pain originating from the insertion of large \textcolor{black}{(18 gauge)} needles. In addition, CP was shown to have a low specificity \textcolor{black}{of only 65\% for ACS when $\Delta$p$<$30 mmHg} \cite{Nelson2013-bo}. Thus, fasciotomies are often carried out prophylactically \cite{Gordon2018-cx} leading to unnecessary trauma, emphasizing the need for new diagnostic tools. 

\citeauthor{McMillan2019-bu} summarized the many technologies that aim to improve the diagnosis of ACS and are currently under investigation, including monitoring localized oxygenation, monitoring of localized perfusion, localized metabolic analysis (glucose, pH), and systemic physiology based \textcolor{black}{on serum biomarkers} \cite{McMillan2019-bu}. 
\textcolor{black}{Among all mentioned methods, monitoring of localized oxygenation attracted considerable attention. In principle, two variables are interesting: oxygen tension, i.e. partial pressure of oxygen (pO$_2$) within the interstitial space and oxygen saturation, i.e. the fraction of hemoglobin that carries oxygen relative to the total hemoglobin in the blood. To measure oxygen saturation, non-invasive near-infrared spectroscopy is widely used and was evaluated for ACS but was found to suffer severe limitations due to the small penetration depth and adverse influence from changes in skin color. \cite{De_Santis2015-fq,McMillan2019-bu}. On the other side, monitoring of pO$_2$} was \textcolor{black}{found} to have certain advantages over measuring pressure in the diagnosis of compartment syndrome in mice \cite{Seekamp1998-oi} as well as canine models \cite{Doro2014-gu,Weick2016-zv}. The canine models studied by \citeauthor{Doro2014-gu} have shown that measuring pO$_2$ had a high specificity and sensitivity for the diagnosis of compartment syndrome \cite{Doro2014-gu}. One clinical study \cite{Hansen2013-yd} evaluated intramuscular oxygen measurements in patients with tibia fractures and found that pO$_2$ could be a good metric to reduce the number of unnecessary fasciotomies. Human data \textcolor{black}{are} very limited due to the non-availability of suitable clinical intramuscular oxygen probes. The only probes currently approved (not for intramuscular measurements) and available for clinical use are Clark-type electrodes, \cite{Clark1953-se} which are severely limited by their long warm-up times (which is not suitable for an emergency setting) as well as frequent re-calibration, while they are also incredibly fragile, underscoring the urgent need for new clinical intramuscular oxygen probes.

Oxygen tension within tissue can be measured using a method known as phosphorescence quenching, where the collision of oxygen with specific phosphorescent molecules can be used to quantify oxygen concentration \cite{Vanderkooi1987-at,Rumsey1988-jn,Dunphy2002-fx, Vinogradov1996-ei, Vinogradov1999-mv,Stich2010-om,Wang2014-pi}. Numerous oxygen sensing molecules have been synthesized, with porphyrins being particularly useful for the measurement and imaging of tissue oxygen tension \cite{Roussakis2015-lv}. Brightly emitting metalloporphyrin oxygen sensors have recently been synthesized that offer high-sensitivity oxygen tension measurements. \cite{Roussakis2015-vu} When excited with blue ($\lambda=377$~nm) or green ($\lambda=531$~nm) light, these Pt(II)-core porphyrins exhibit red phosphorescence ($\lambda=645$~nm) which is inversely proportional to pO$_2$ according to the Stern-Volmer relation \cite{Stern1919-mp}:

\begin{equation}
    I = \frac{I_{0}}{1+k\cdot[pO_{2}]}
    \label{eq:SV1D}
\end{equation}

where $k$ is the Stern-Volmer quenching constant and I$_0$ the intensity of the phosphorescence in the absence of a quencher (oxygen). The bright red light from these new porphyrins can be seen by the naked eye and quantified with the help of portable imaging equipment.  These new porphyrins have been clinically validated as part of liquid bandages for the assessment of wound healing \cite{Koolen2017-rt,Li2018-mh} and integrated into wearable devices for performance monitoring \cite{Roussakis2020-ks}.

The further development of the above-mentioned portable technology into a toolkit for sensing the loss of deep tissue oxygenation associated with compartment syndrome is described herein. This was accomplished via the integration of an oxygen-sensing material with optical fibers and hypodermic needles or catheters. 

\textcolor{black}{
Fiber-optic oxygen sensors based on phosphorescence quenching are well described in scientific literature \cite{Sayuri-ls,Chen2014-oo,Davenport2016-fk,Kocincova2007-fs,Wolfbeis2015-qp,Chen2013-vt,Chen2016-ri,Formenti2015-od,Chen2012-qm} and have been evaluated in \textit{in vivo} models \cite{Weiss1997-cr,Ye2007-lg,Yin2009-rp,Al-Mutawa2018-lr,Crockett2019-iy,Jordan2004-kf,Mahling2015-am}. However, to date, there is no medical device available for the clinical measurement of intramuscular oxygen. This might be due to the fact that most existing sensors were not designed to measure oxygen under physiological conditions in a clinical setting, require expensive and oversized readout devices, while many of these sensors have also not been evaluated beyond limited \textit{in vivo} models. For example, the OXY-MICRO-AOT from World Precision Instruments was evaluated in surgically exposed epididymal fat pads \cite{Ye2007-lg,Yin2009-rp}, whereas the oxygen microsensor from PreSens was evaluated in tumors on the chorioallantoic membrane of chick embryos \cite{Al-Mutawa2018-lr}. Also, intravascular sensors such as the discontinued Paratrend \cite{Weiss1997-cr} are not directly applicable to an intramuscular oxygen measurement where insertion force is applied to the sensor. A luminescence based sensor which has been extensively used in various \textit{in vivo} applications, \cite{Driessen2007-kg,Jordan2004-kf,Mahling2015-am} even for intramuscular measurements, is the OxyLite from OXFORD OPTRONIX, however, only their system WellBeing is currently being evaluated in a clinical trial for the purpose of bladder tissue oxygen measurement and the readout device is rather large.}

The oxygen sensing device described here could be especially useful to assess not only ACS, but also other pathological conditions, such as vascular diseases, diabetic wounds, burns, cancer, and traumatic injuries, that can result in a reduction of tissue pO$_2$ leading to hypoxia \cite{Carreau2011-qv}.

\section{EXPERIMENTAL SECTION}
\subsection{Materials.}
In order to synthesize an optimally-performing material for optical fiber-based deep tissue oxygen sensing, a variety of compounds and formulations were tested. The primary goal was to find a biocompatible host matrix material that would also be chemically compatible with the metalloporphyrin molecules so that aggregation can be avoided. The resulting oxygen-sensing material would need to display high pO$_2$ sensitivity in the physiological range of 0-80~mmHg under humid conditions while being insensitive to changes in pH. Furthermore, the material was required to adhere well to the tip of small-diameter optical fibers. Developing a material and coating process compatible with small fiber diameters was critical, as the goal was to ultimately use needles of the smallest gauge possible to limit patient discomfort. In addition, it was important to keep the fiber pre-processing as simple as possible to facilitate the rapid translation of the device to both military and civilian patients. For this purpose, five different matrix materials were investigated: Tetraethyl Orthosilicate (TEOS) \textcolor{black}{sol-gel}, \textcolor{black}{3M Cavilon}, Poly(ethyl methacrylate) (PEMA) and Poly(propyl methacrylate) (PPMA).

Due to its noted compatibility with porphyrins and the humidity-insensitivity described in Ref.~\citenum{Roussakis2020-ks} tetraethyl orthosilicate (TEOS) \textcolor{black}{sol-gel} was a strong candidate material. TEOS-containing matrices have been used before to produce spin-coated \cite{Tang2003-qi,Ismail2013-wd} as well as fiber-based oxygen sensors \cite{Jorge2004-ub,Yeh2006-hy,Lo2008-rn,Chu2007-ch,Chu2011-go,Chu2012-ho} using commercially available ruthenium and platinum complexes. However, the fibers used in these references all had large diameters of $\geq$550~$\mu$m incompatable with small bore needles, or the fiber tip was further processed for example by tapering to increase the signal strength. \textcolor{black}{In addition, ruthenium complexes are not suited for \textit{in vivo} applications due to their toxicity.}

For the current work, TEOS formulations containing 50~$\mu$M of the alkyne-terminated Pt(II) porphyrin previously developed in-house were prepared in a similar fashion to the procedure described in Ref.~\citenum{Roussakis2020-ks}. 
\textcolor{black}{In-house synthesized porphyrin derivatives were chosen for this work due to our familiarity with their properties and their performance across different materials.  Furthermore, we have established  synthetic  protocols  for their  derivatization  that  can  be  used  in  future  immobilization within matrix materials, either by chemical attachment or photo-crosslinking.}
The alkyne-terminated porphyrin\textcolor{black}{, whose molecular structure is shown in Fig.~\ref{fig:porphyrins},} was synthesized as described in Ref.~\citenum{Roussakis2015-vu}. TEOS, 1-octanoyl-rac-glycerol (referred to as polyol) and dimethyl sulfoxide (DMSO) were purchased from Sigma-Aldrich. Ethyl alcohol and hydrochloric acid were purchased from Fisher Scientific. For a final, 50 $\mu$l TEOS/porphyrin formulation at pH~1, a solution of TEOS and a solution of the polyol and the porphyrin were prepared in two separate, small Eppendorf tubes. In the first tube, 12.5~$\mu$l TEOS (25~wt\%) was added to 15~$\mu$l DMSO followed by addition of 1.7~$\mu$l from a 1M hydrochloric acid in ethanol solution. \textcolor{black}{In contrast to the formulations in Ref.~\citenum{Roussakis2020-ks},} here DMSO was used instead of ethanol to increase the surface tension. In the second tube, 5 mg of the 1-octanoyl-rac-glycerol (10 wt\% polyol) was mixed with an aliquot of the alkynyl metalloporphyrin stock solution in dichloromethane (DCM) and ethanol was added to bring the combined volume of both tubes to 50~$\mu$l. After vortexing, the polyol solution was added to the TEOS tube, vortexed again, and was left to sit for 15-20 minutes. In order to reduce cracking, additional formulations were made where surfactants of different amounts (1.4~wt\% and 2.8~wt\% of Triton X-100 and Tween-20) were added, similar to approaches described in Refs.~\citenum{Ismail2013-wd,Yeh2006-hy,Lo2008-rn}.

The second material investigated for the purpose of coating optical fibers was the terpolymer-based \emph{\textcolor{black}{3M Cavilon} No Sting Barrier Film}. \textcolor{black}{3M Cavilon} is an FDA approved liquid bandage, is water-resistant, adheres well to various surfaces, and is breathable. Since cavilon is hydrophobic it was found immiscible with the alkyne-terminated porphyrin. Instead, the much more hydrophobic pivaloyl-terminated derivative was used, which displayed good miscibility. The \textcolor{black}{Pt(II)-pivaloyl}-terminated porphyrin was synthesized as described in Ref.~\citenum{Roussakis2015-vu}. For the final solution, \textcolor{black}{3M Cavilon} was vortexed with the pivaloyl-terminated porphyrin in an Eppendorf tube.

\textcolor{black}{
\begin{figure}[tb]
\centering
\includegraphics[width=\linewidth]{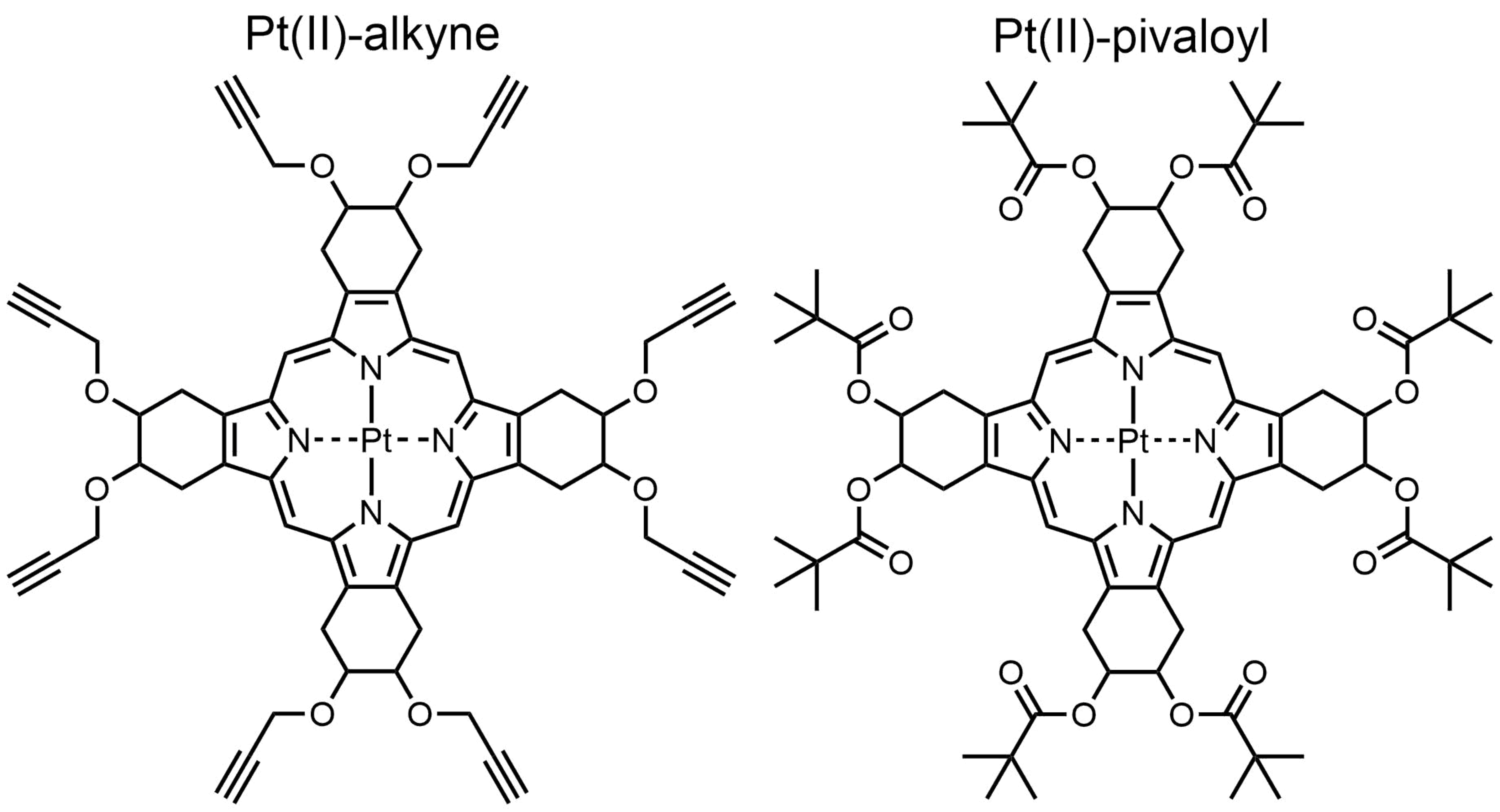}
\caption{\textcolor{black}{Molecular structure of the Pt(II)-alkyne porphyrin (left) and the Pt(II)-pivaloyl porphyrin (right) used for the formulations.}}
\label{fig:porphyrins}
\end{figure}}

In addition to \textcolor{black}{TEOS sol-gel} and cavilon, PEMA and PPMA were investigated for use with the pivaloyl-terminated porphyrin. These acrylate polymers together with poly(methyl methacrylate) (PMMA) were investigated previously by \citeauthor{Chen2014-oo}  for the purpose of coating the tapered tip of 480~$\mu$m  fiber oxygen sensors \cite{Chen2014-oo}. They showed that PPMA had higher sensitivity and faster response time than PEMA and PMMA. \textcolor{black}{PEMA was used before by \citeauthor{Chen2012-qm} to coat 200~$\mu$m optical fibers \cite{Chen2012-qm}.} \citeauthor{Davenport2016-fk} additionally combined a PtOEP phosphor with PEMA and coated 600~$\mu$m optical fibers \cite{Davenport2016-fk}. 

For coating the 200~$\mu$m fibers to produce the sensor described in this publication, a solution was made as follows: in an Eppendorf tube 0.025~mg/$\mu$l PEMA or PPMA (purchased from Sigma-Aldrich and Scientific Polymer Products) was dissolved in DCM. After vortexing, the pivaloyl terminated porphyrin was added at a concentration of 50~$\mu$M and the solution was vortexed again. While higher and lower concentrations of PEMA and PPMA were also tested, the coating materials made from solutions at a concentration of 0.025~mg/$\mu$l were found to display the largest oxygen-sensing response.

In order to prevent the direct contact of the oxygen sensing material with bodily fluids and to shield the material from external light, all fiber tips were additionally coated with a breathable, reflective white layer. The coating had the advantage of increasing the measured porphyrin emission signal due to back-reflection and protecting the oxygen sensing layer during a future sterilization process using ethylene oxide (EtO). EtO sterilization is used as the standard method to sterilize biomedical sensors since it has been shown to have no damaging effects on polymer coatings and silica fibers \cite{Zhang1991-md,Stolov2013-fq}. Similar to the coating described in Ref.~\citenum{Hahn1994-lf}, the protective coating for the oxygen sensor contained a white pigment based on 40\% titanium dioxide and silicone. To prepare it, 0.1~g of a \textcolor{black}{dimethylsiloxane} copolymer (Gelest, CAS 68037-59-2) and 1~g of white pigment concentrate (Gelest, PGWHT01) were mixed together in an Eppendorf tube. Subsequently, approximately 0.03~g of cure retarder (Gelest, Utensil R1) and a small drop of platinum catalyst (Sigma Aldrich, CAS 68478-92-2) were added and the solution was stirred vigorously. The combination of the cure retarder and catalyst resulted in a mixture that lasted for approximately 10 minutes before solidifying, providing ample time to coat the fiber tips.

\subsection{Fabrication of the fiber sensors.}
Multimode silica fibers with a 200~$\mu$m core (Thorlabs, FP200URT) were cleaned with isopropyl alcohol, and the ends were stripped and cleaved. 

For the TEOS-based coating, the fiber tip was functionalized either by plasma treatment (BD-20AC plasma treater) or by silanization (1~wt\%-Aminopropyl triethoxysilane in water solution) immediately before coating to improve the adherence of the \textcolor{black}{TEOS sol-gel} layer. The fibers were dipped in the silane solution for 10 minutes and were then dried at 120$^{\circ}$C for a period of 2 hours. Subsequently, the fiber tips were dipped (1-5 times) by hand in the TEOS solution. Dipping the fiber several times increased the thickness of the layer and thus improved the signal strength. After dipping, the fibers were left to dry overnight and were then placed under high vacuum for 2-3 hours the next morning. 

For the cavilon formulation as well as the PEMA, and PPMA matrices, the fibers were not functionalized before coating. A drop of porphyrin-matrix solution was transferred from the Eppendorf tube to the surface of a PDMS film and the fibers were dipped by hand in the porphyrin-matrix solution. The fibers were left to dry overnight and were subsequently placed under high vacuum for 2-3 hours.

\begin{figure*}[ht]
\centering
\includegraphics[width=\linewidth]{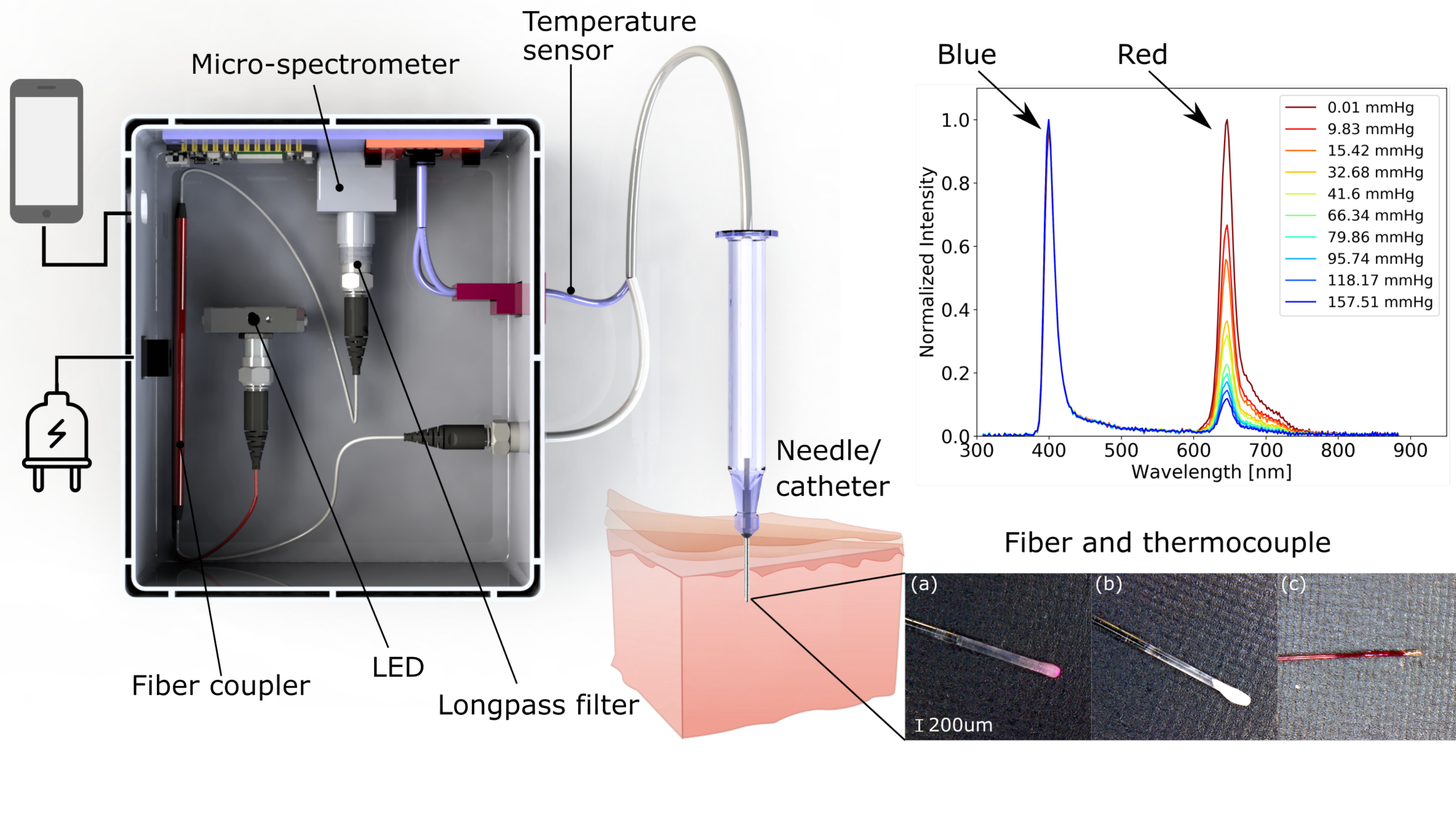}
\caption{\textcolor{black}{Prototype device (10x11~cm, weighting 200~g) showing the (a) needle/catheter containing the porphyrin-PPMA coated fiber, (b) the needle/catheter after recoating with silicone, and (c) a thermocouple which was placed inside the needle next to the fiber. The obtained spectrum is shown on the right with emission (red) and excitation peak (blue). The 375 nm excitation light was partially blocked with a 400 nm longpass filter, with the peak seen in the spectrum being the tail of the excitation light from the LED which was not blocked by the filter.}}
\label{fig:Box}
\end{figure*}

After coating with the oxygen sensing layer, the fibers were dipped in the white silicone coating, then dried with 100$^{\circ}$C hot air for 10-15 seconds, and were finally left to dry for 48 hours at room air. To strengthen the adherence of the silicone layer and prevent it from detaching from the fiber, an additional protective layer of \textcolor{black}{3M Cavilon} was coated on the fiber.

\subsection{Apparatus and data acquisition.}
Optical and electronic hardware was fitted inside a 10x11x4.5~cm 3D printed box  as shown in Figure~\ref{fig:Box}. A Hamamatsu C12880MA-10 micro-spectrometer was connected to a custom printed circuit board (PCB) for read-out and for driving the LED source (Thorlabs, LED375L). The PCB was connected to a Particle Photon microcontroller and an external 12~V power supply to provide stable voltage. A solely USB-based power supply was found to not be stable for both pulsing the LED and reading out the micro spectrometer simultaneously. The firmware for the photon microcontroller was custom-built on the base of Arduino example code from Ref.~\citenum{Groupgetsgit}.

The 375~nm excitation light as well as the 650~nm phosphorescence light was guided through a 1x2 fiber optic coupler (Thorlabs TH200R5S1B). On the micro-spectrometer side of the coupler, a piece of a flexible UV filter from Edmund optics ($\#$39-426, \textcolor{black}{400 nm longpass}) was glued to the SMA connector in order to reduce the contribution from the blue light which was otherwise saturating the spectrometer. An SMA-to-FC/PC mating sleeve (Thorlabs ADAFCSMA1) was mounted in the wall of the box where the fiber optic sensor was connected.

The base for the fiber-optic sensor was an FC/PC connectorized custom fiber optic fiber patch cable from Thorlabs with a 200~$\mu$m core  and 0.5~NA (FP200URT). The patch cable was cut, the tubing removed from one side and prepared and coated as described earlier.

\begin{figure}[ht]
\centering
\includegraphics[width=\linewidth]{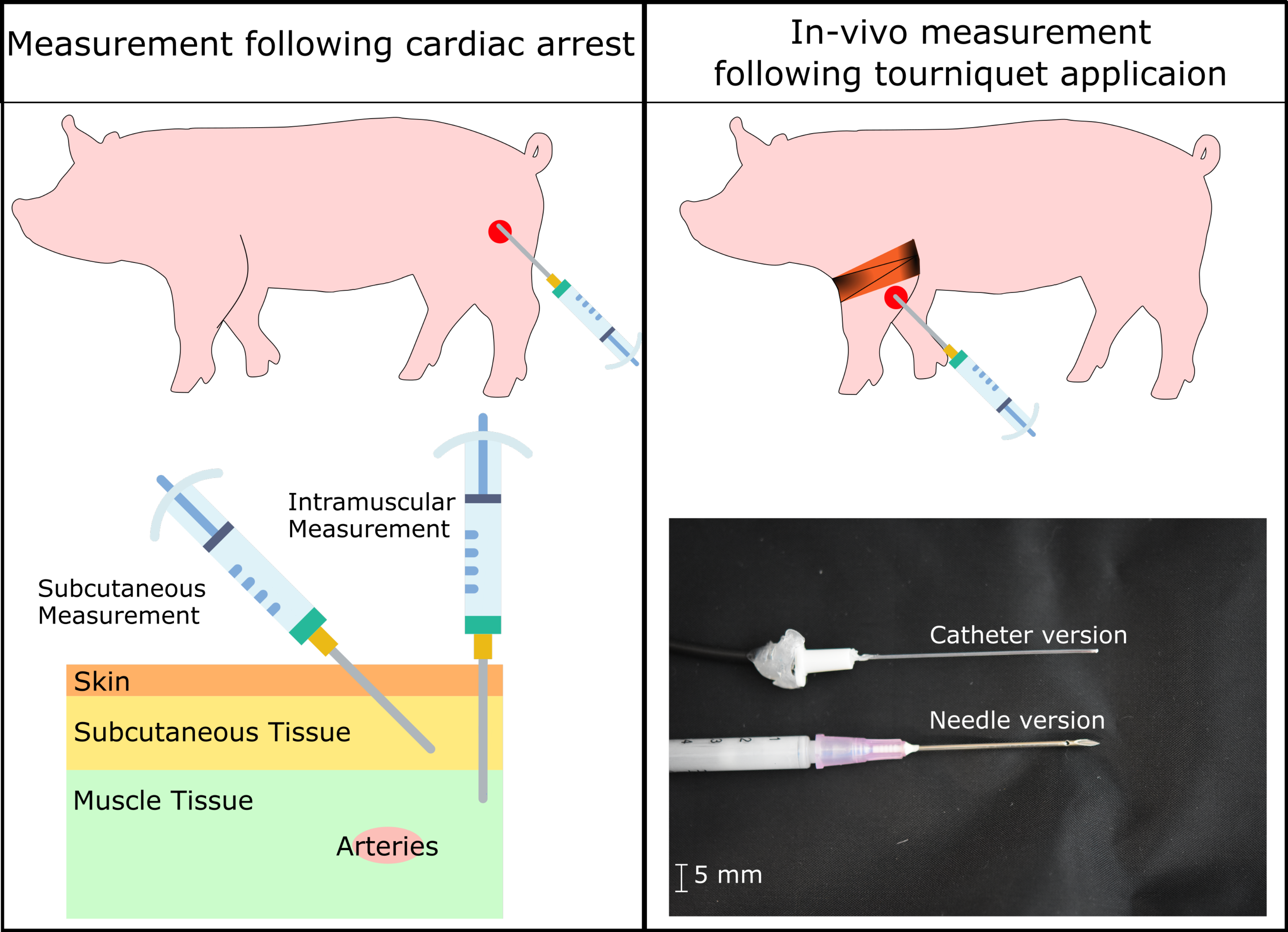}
\caption{\textcolor{black}{Experiments in a porcine model. In the first experiment, limb oxygenation was measured following loss of cardiac function in a Yorkshire swine. The oxygen measurement was carried out at the hind limb (biceps femoris muscle). The needle was subsequently inserted into muscle tissue, the subcutaneous tissue, and again into muscle tissue. In the second experiment, limb oxygenation following the placement of a tourniquet was measured in two pigs. The oxygen sensor was inserted into the flexor carpi ulnaris muscle in the front limb. The tourniquet was applied above the elbow joint over the Tricheps Brachii and Brachialis muscles.}}
\label{fig:PigMethods}
\end{figure}

Together with a 24 gauge thermocouple (IT-24P, physitemp), the coated fiber was integrated into two different versions of the device. In the first version, the fiber was glued into a 18 gauge needle (\textcolor{black}{BD PrecisionGlide})\textcolor{black}{, shown in Figure~\ref{fig:PigMethods}}. To allow for adequate equilibration with the surrounding tissue, two 1~mm side ports were drilled into the fiber 5~mm above the tip. The holes were at the same height and separated by 180 degrees. To prevent the fiber tip from breaking when inserted into tissue, the needle tip was closed using light cure medical device adhesive (Loctite 3321). A 1 ml syringe body (\textcolor{black}{HSW Norm-Ject Tuberkulin}) was used to provide stability and easier handling to the fiber sensor. It should be noted that the 200~$\mu$m fiber could fit in a smaller gauge needle in the future and the 18 gauge needle was only selected for accuracy in creating the side-port holes.

In the second version of the device, the fiber was integrated into a flexible polyethylene tubing with an outer diameter of 0.6 mm and a luer connector was added to ensure tight mounting in a standard catheter. The length of the tubing was chosen in a way that the oxygen sensing part would peek out of the tip of a standard 20 gauge catheter (Exel Safelet catheter, 20G x 1 1/4'') when the luer was locked in place.   

The LED pulsing time as well as the time between measurements were adjusted to the signal intensity of the different coatings using software settings. For the final device, a pulsing time of 5~ms and measurement interval of 15 seconds were used. The microcontroller was read out via a USB cable using a smartphone. A USB connection was chosen over wireless connection to add an additional layer on data safety when dealing with clinical data. An android smartphone application was developed with the help of Google's Flutter software development kit (SDK) and Android Studio. The application provided the option to change the settings such as pulsing time, measuring interval, output file name, and pO$_2$ calibration. Besides the current spectra, the application displayed the pO$_2$ timeline and provided the option to store the data file in a text format on the SD card of the smartphone.

For calibration and testing purposes, the oxygen sensor was placed in a small gas chamber alongside a commercial laboratory oxygen sensor (Profiling Oxygen Microsensor PM-PSt7, PreSens) which provided an independent readout of the chamber pO$_2$. The temperature of the gas chamber was adjusted with the help of a hot plate. The oxygen partial pressure in the chamber was adjusted between 0~mmHg  and 160~mmHg by changing the relative flows of nitrogen and air with the help of a gas mixer. The humidifier allowed to switch between dry and humid conditions.

To extract pO$_{2}$ from the data, a non-linear least squares fit was used, which was based on a two-dimensional Stern-Volmer relation containing a linear dependence on temperature:

\begin{equation}
    I = \frac{I_{0}}{1+(k_{0}+k_{T}\cdot(T-T_{C}))\cdot[pO_{2}]} + f 
\label{eq:SV2D}
\end{equation}

where $f$ accounts for the phosphorescence from the porphyrin molecules non-accessible by quenchers \cite{Lakowicz2006-ef}, $k_T$ is the temperature dependent quenching constant, and $T_C$ the room temperature at which the calibration was performed. The intensity $I$ was extracted from the spectrum by integrating the red spectral range and normalizing it to the blue excitation light. The linear temperature dependence was verified by sweeping through temperatures at fixed values of pO$_{2}$ (\textcolor{black}{Supporting Figure ~S1}). pO$_2$ as well as temperature were changed in both directions and no significant hysteresis effects were observed.

\begin{figure}[ht]
\centering
\includegraphics[width=\linewidth]{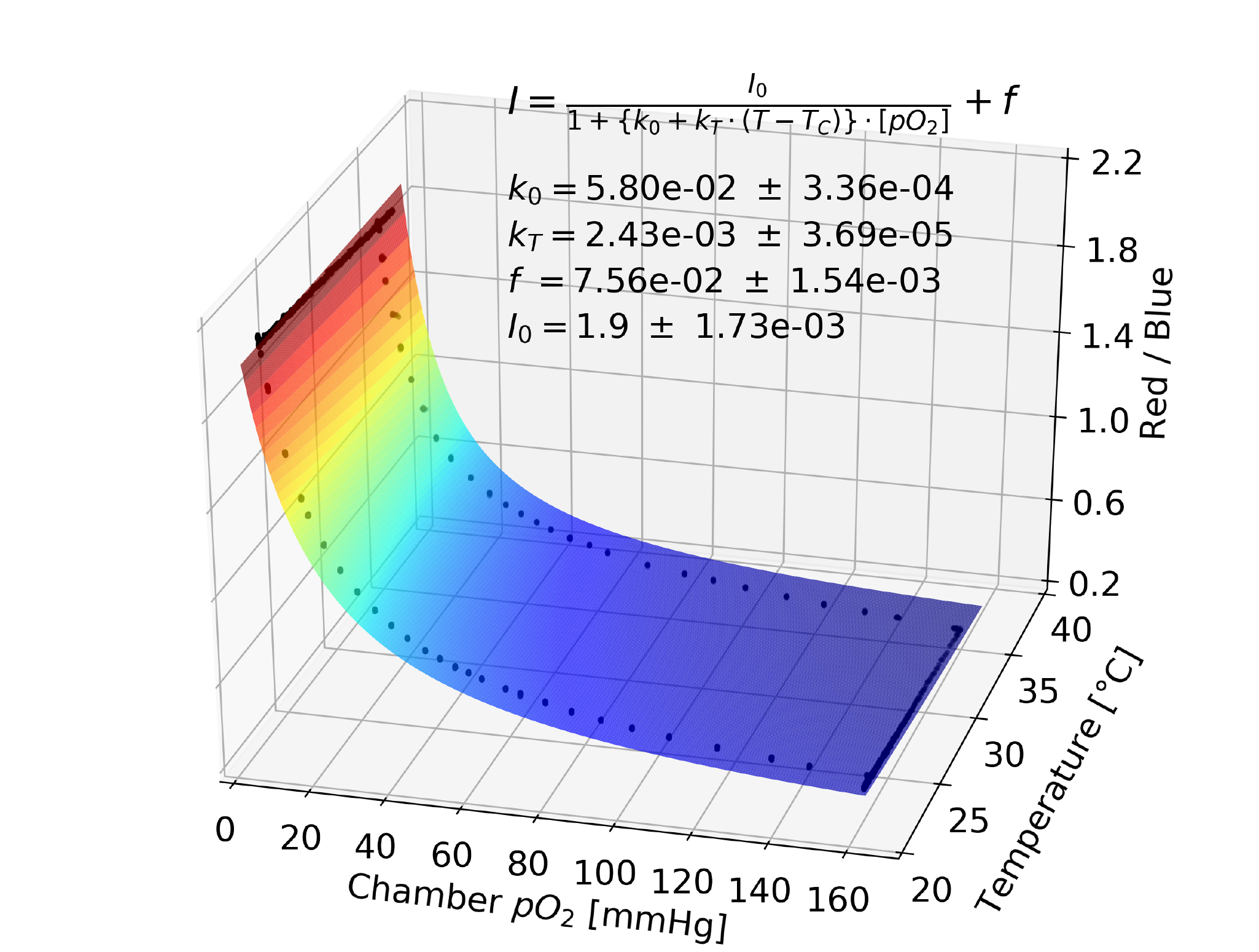}
\caption{2D calibration obtained from a \textcolor{black}{Pt(II)-pivaloyl in PPMA} coated fiber. \textcolor{black}{$I$ is the intensity of the phosphorescence (Red/Blue), $I_0$ is the phosphorescence at zero oxygen, T is the temperature, $T_C$ is the calibration temperature from the 1D calibration, $k_0$ and $k_T$ are the temperature independent and temperature dependent Stern-Volmer quenching constants, respectively, and $f$ accounts for  the  phosphorescence  from  the porphyrin molecules non-accessible by quenchers.}}
\label{fig:2DCalib}
\end{figure}

The resulting two-dimensional calibration plot for a fiber coated with the PPMA matrix in shown in Figure~\ref{fig:2DCalib} \textcolor{black}{and the calibration parameters for all three fiber sensors used in the porcine studies are given in Supporting Table~S2}. By extracting the fit parameters $I_{0}$, $k_{0}$, $k_{T}$ and knowing the calibration temperature $T_{C}$, pO$_2$ was deduced from Eq.~\ref{eq:SV2D}. The corresponding pO$_2$ errors were calculated from the same equation using Gaussian error propagation and the parameter fitting errors. The error for the laboratory oxygen sensor was 3\% and the error of the temperature sensor was 1$^{\circ}$C. 

The resulting pO$_2$ response was matching the design requirements in the physiological range between 0 and 80~mmHg where errors were smaller than 5\%. At higher pO$_2$ values, the errors increased to approximately 6\% \textcolor{black}{(Supporting Figure~S2)}. This was expected, as the porphyrin emission intensity decreases with increasing pO$_2$. In the future, this could be improved by adding a second porphyrin or porphyrin-containing material which is tuned to be more sensitive at higher values of pO$_2$.

\textcolor{black}{
\subsection{Porcine model.}}

To assess the performance of the sensor under realistic conditions, two sets of \textit{in vivo} experiments were performed in two Yorkshire swines and one Hampshire swine, with all swine being female. The similar anatomic scales between swine and humans makes the porcine model an especially suitable model to study physiological changes in intramuscular oxygenation. The animal protocol was reviewed and approved by the Institutional Animal Care and Use Committee at Massachusetts General Hospital, and all procedures were performed within the Knight Surgical Facility. \textcolor{black}{The conducted study was a pilot study and was meant to gather meaningful, but not statistically significant data.}
For all pigs, anaesthesia was induced with intramuscular Telazol (4.4~mg/kg) and Atropine (0.4~mg/kg) followed by an Isoflurane (1-3\%) inhalation. During the entire procedure, the pigs were ventilated with a Fraction of Inspired Oxygen (FiO$_2$) of 1 \textcolor{black}{which is the standard procedure for pigs at Massachusetts General Hospital. In previous experiments, ventilation at lower FiO$_2$ was seen to lead to a decrease in blood oxygen saturation due to shallow breathing under anaesthesia of these animals. In order to prevent hypoxia related conditions, the standard protocol at FiO$_2$=1 was used. The pigs for the current study were an on-table transfer from another protocol and had undergone previous procedures related to different laser skin treatments.}

In the first experiment, limb oxygenation following loss of cardiac function was measured in a pig. The oxygen sensing needle prototype was sequentially inserted into several regions of the pig's hind limb at the biceps femoris muscle one minute following euthanasia via administering Pentobarbital euthanasia solution (Fatal Plus) \textcolor{black}{as shown in Figure~\ref{fig:PigMethods}}.

In the second \textit{in vivo} experiment limb oxygenation following the placement of a tourniquet was measured in two pigs. A tourniquet was applied for 30 minutes to the front limb of two pigs above the elbow joint over the Tricheps Brachii and Brachialis muscles, as shown in Figure~\ref{fig:PigMethods}. \textcolor{black}{Due to the conical shape and the shortness of the pig leg, a standard pressurized tourniquet could not be used, instead a RATS GEN 2 tourniquet was applied over a rubber tourniquet (SWAT-T) and tightened by hand.}

Before applying the tourniquet, the oxygen sensor was inserted into the flexor carpi ulnaris muscle (indicated by the red dot in Figure \ref{fig:PigMethods}). In one pig, the oxygen sensor was needle-based whereas in the second pig the catheter version \textcolor{black}{(shown in Fig.~\ref{fig:PigMethods} on the lower right)} of the oxygen sensor was used. In addition to the prototype oxygen sensor, a laboratory-grade oxygen sensor (Profiling Oxygen Microsensor PM-PSt7, PreSens) was  inserted with a catheter nearby in both experiments. \textcolor{black}{All oxygenation measurements were temperature compensated using the included thermocouple.}
\vspace{5mm}

\section{RESULTS AND DISCUSSIONS}
\subsection{Oxygen response.}
\begin{figure}[bth]
\centering
\includegraphics[width=\linewidth]{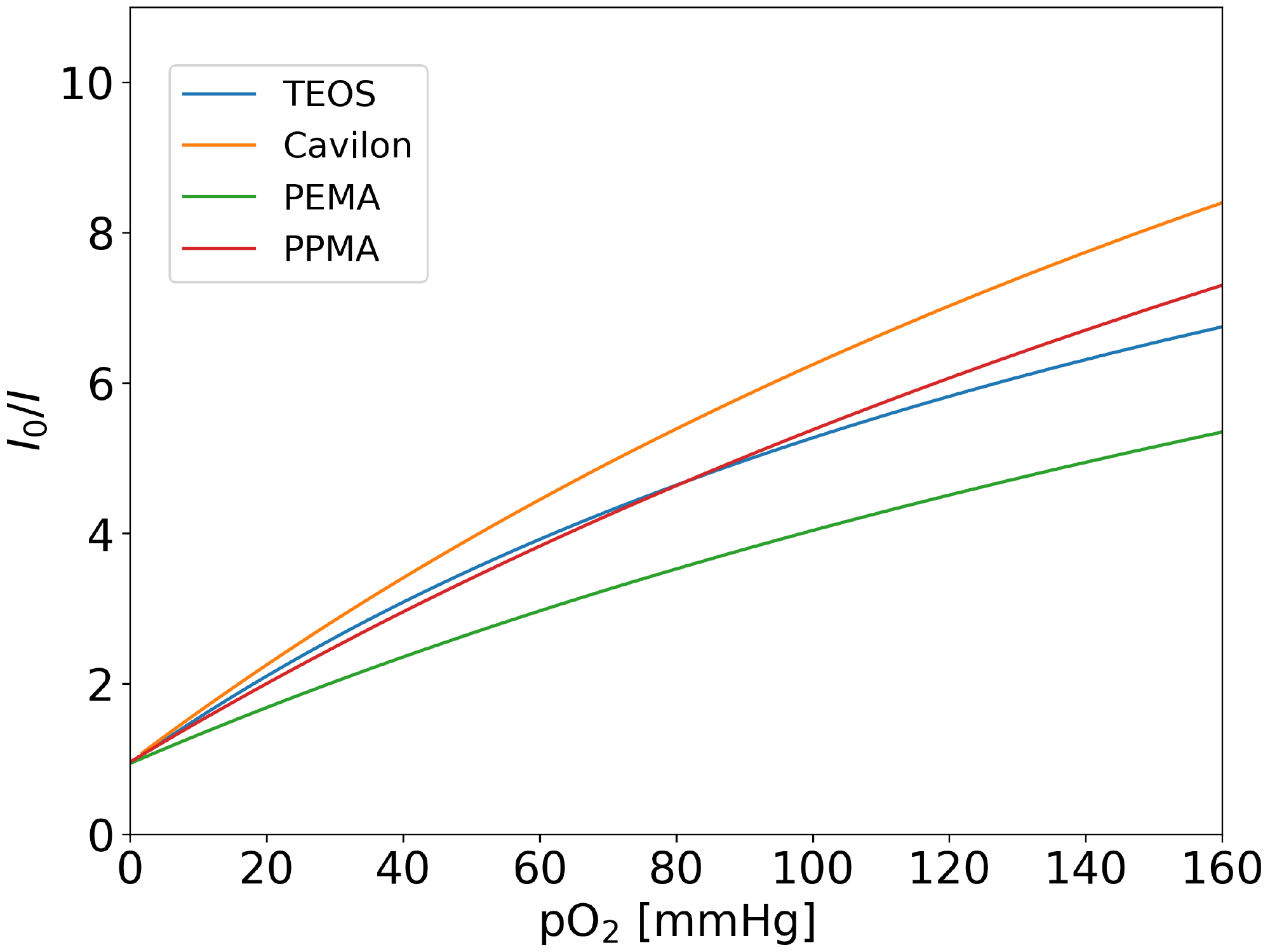}
\caption{Comparison of the Stern-Volmer distributions extracted for different matrices: \textcolor{black}{50 $\mu$M Pt(II)-pivaloyl in TEOS sol-gel + 1.4~wt\% Triton X-100, 50 $\mu$M Pt(II)-pivaloyl in Cavilon, in 0.025~mg/$\mu$l PEMA, and in 0.025~mg/$\mu$l PPMA.}}
\label{fig:Compare1DSternVolmer}
\end{figure}

For the material selection process, the intensity of phosphorescence signal was measured for different levels of pO$_2$ at room temperature and fitted with a one-dimensional Stern-Volmer relation as in Eq.~\ref{eq:SV1D} modified with a factor $f$ for non-accessible molecules. The resulting Stern-Volmer distributions are shown in Figure~\ref{fig:Compare1DSternVolmer} and the corresponding calibration curves are shown in Supporting Figure~S3. As can be clearly seen, the pivaloyl porphyrin in the cavilon matrix was the most sensitive over the whole measured pO$_{2}$ range with $k=0.0742$, $f=0.0893$ and $I_0=2.2$. The PEMA coating was the least sensitive with $k=0.0442$, $f=0.0771$ and $I_0=1.2$. The sensitivities of the \textcolor{black}{TEOS sol-gel} and PPMA were comparable; the \textcolor{black}{TEOS sol-gel} was slightly better at very low pO$_{2}$ and PPMA was better at high pO$_{2}$. The fitting parameters for PPMA were $k=0.0593$, $f=0.0784$ and $I_0=1.9$, while for \textcolor{black}{TEOS sol-gel} they were $k=0.0731$, $f=0.16$ and $I_0=2.2$. Hence, the \textcolor{black}{TEOS sol-gel} matrix had the highest fraction of inaccessible porphyrins. The coating quality of the \textcolor{black}{TEOS sol-gel} material was varying considerably between different fibers and in general the \textcolor{black}{TEOS sol-gel} did not adhere well, which can be attributed to significant cracking and had been already observed in prior studies \cite{Ismail2013-wd,Tang2003-qi}. Applying only minimal force to the fiber tip resulted in a loss of the oxygen sensing layer. Adding surfactants yielded higher emission signals (best results with \textcolor{black}{TEOS sol-gel} + 1.4~wt\% Triton X-100, shown in \textcolor{black}{Supporting Figure~S4}), however the \textcolor{black}{TEOS sol-gel} still did not adhere well to the fiber tip.

\subsection{Humidity sensitivity.}
During its application the oxygen sensor will be exposed to blood and other bodily fluids, thus humidity insensitivity is crucial.

The humidity sensitivity was measured by calculating the ratio of the signal in the emission peak under dry and humid ($>$90\%) conditions at 0~mmHg, the corresponding spectra are shown in Supporting Figure~S5. The pure \textcolor{black}{TEOS sol-gel} matrix was seen to be humidity sensitive with a dry-to-humid ratio of 2.1. \textcolor{black}{While TEOS-based sol-gels have been reported earlier that did not display humidity sensitivity \cite{Roussakis2020-ks}, the materials used in this work were made from formulations that contained DMSO and surfactants. Additional measurements on fibers coated with TEOS sol-gels of different compositions (Supporting Figure S5) have shown that the humidity sensitivity on the fibers is mainly a result of the addition of DMSO to the TEOS formulation. The uniformity of the coating on a support substrate also plays a role, as the addition of surfactants has been found to improve performance by minimizing the formation of cracks in the sol-gels coated on fibers.}

The humidity sensitivity of TEOS decreased when surfactants were added, reaching a dry-to-humid ratio of 1.3 to 2. This likely arised as pure TEOS coatings showed significantly more cracks and thus provided more possibilities for water to enter the coating

No acrylate polymer\textcolor{black}{-}based coating showed significant humidity sensitivity, which can be attributed to their hydrophobicity\textcolor{black}{, the values are given in Supporting Table S1}. 

\subsection{Photobleaching.}
The intensity of phosphorescence emission from the oxygen sensing materials was measured over time to determine the photobleaching rate. For this set of experiments, the LED was set to pulse every 15 seconds for a total duration of 1.5 hours. The irradiance from the LED on the oxygen sensing material was estimated to be 160~nW/cm$^2$. 

\textcolor{black}{TEOS sol-gel} was found to have the highest bleaching rate with 5.0\% per hour, followed by PEMA with 2.9\% and cavilon with 1.8\%. PPMA was observed to have the lowest bleaching rate with only 0.2\%. As the sensing tool required for compartment syndrome would be used only once for a duration up to 10 hours, this minimal level of photobleaching would lead to a minor change in overall brightness, ensuring high accuracy during use. Moreover, by carefully counting the number of total delivered LED pulses, the small photobleaching rate was calibrated and accounted for. 

Given its low photobleaching rate and humidity insensitivity, PPMA was selected as the optimal matrix material for the deep tissue oxygen sensor. \textcolor{black}{The pivaloyl porphyrin within the PPMA matrix was further assessed by acquiring phosphorescence intensity spectra and lifetime decays (Supporting Figure~S6), using the FLS1000 steady state and phosphorescence lifetime spectrometer by Edinburgh Instruments (Livingston, UK). The resulting lifetime within the PPMA matrix was determined to be 98 $\mu$s which is almost identical to the lifetime in DCM ($\tau_0=101$ $\mu$s) indicating good compatibility of the oxygen sensing molecule and the matrix material.}

\subsection{pH sensitivity.}
The material with \textcolor{black}{the} lowest photobleaching rate and humidity insensitivity, PPMA was investigated for pH sensitivity. This was especially important for the application of the sensor: during muscle injuries, the intramuscular pH is expected to decrease from above pH~7 to below pH~5 \cite{Challa2017-mg,Gershuni1985-dd,Raza2015-ey}. Therefore, stability over this pH range is crucial. 

To measure pH sensitivity, the oxygen sensor was immersed into a buffer solution at pH~7.5 at 40~mmHg pO$_2$, with the pH slowly lowered via the addition of 2~M hydrochloric acid drop-wise. An oxygen tension of 40~mmHg was chosen since it is in the middle of the pO$_{2}$ range of interest. No pH dependence could be detected\textcolor{black}{(data shown in Supporting Figure~S7).}

\subsection{Assessing Reusability.}
\begin{figure}[ht]
\centering
\includegraphics[width=0.8\linewidth]{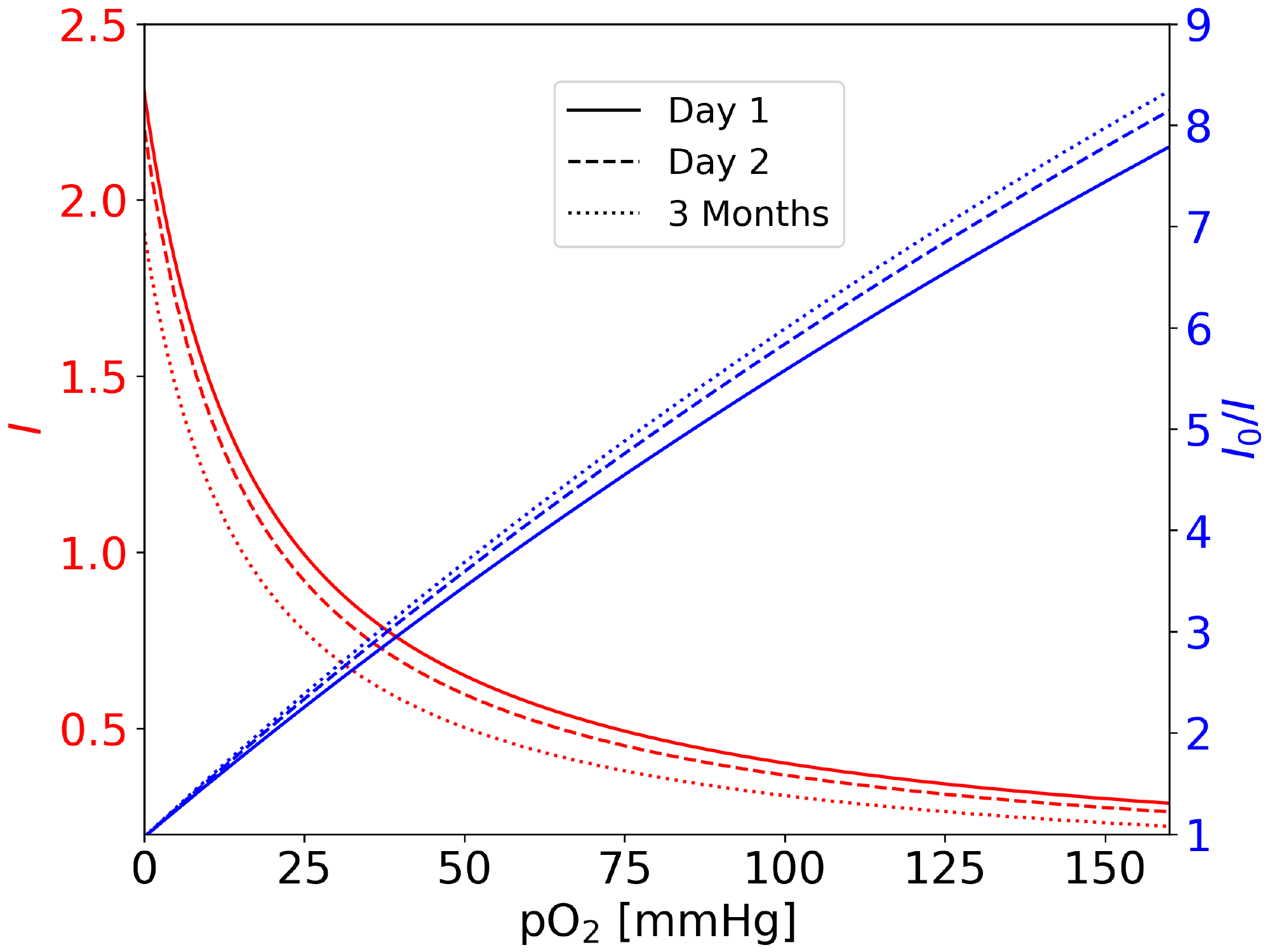}
\caption{Stern-Volmer distributions extracted from the calibration of the same sensor \textcolor{black}{with Pt(II)-pivaloyl-PPMA} coating at different time points \textcolor{black}{and after usage in vivo including cleaning with isopropyl alcohol}. The overall signal intensity $I$ decreased with time whereas the relative sensitivity $I_0/I$ increased.}
\label{fig:ShelfLife}
\end{figure}

All sensors were generally used within 48 hours after they were made, hence ageing of the sensor was not critical for the measurements acquired in this study. To understand the impact of sensor insertion, removal, and cleaning, the sensors were re-evaluated after a period of more than three months from initial use \textit{in vivo} and after rinsing with isopropyl alcohol. \\
The overall signal intensity of the sensors was found to decrease by approximately 30\%, whereas the relative sensitivity increased over time as can be seen in Figure~\ref{fig:ShelfLife}. \textcolor{black}{Further investigations have shown that the drift is only prominent in the measurement of phosphorescence intensity and not in the lifetime measurement, where it was seen to be below 5\% over a period of three months. This indicates that the drift in intensity is originating from changes in the optical properties of the matrix, for example scattering, and not from the oxygen sensing molecule itself. Transferring to a lifetime-based readout in the future will significantly improve potential reusability.} 

\subsection{Response time.}
The final PPMA-based sensor with the silicone coating had a response time of 35~s to reach 1/e when transitioning from 160~mmHg to 0~mmHg. The response time is thought to be mainly limited by the diffusion of oxygen through the silicone layer as well as the physical absorption of oxygen in the silicone layer. It is important to note that this response is more than sufficient for the intended application, as compartment syndrome evolves over 30 minutes to several hours and oxygen levels will change over much smaller intervals. 

\subsection{Leaching studies.}
To assess the biocompatibility of the material and fiber coating, the leaching of the porphyrin molecules from the material was analyzed by measuring the total platinum content found in samples after exposure to coated fibers. For this purpose, fibers were coated with different combinations of porphyrin, PPMA, and silicone recoating using the process described above. After drying, the fibers were placed inside a K2EDTA blood collection tube containing 1~ml of fresh whole pig blood, and kept immersed for a duration of 7 hours. \textcolor{black}{Additionally, fibers in needles were inserted into porcine tissue samples (skin and muscle) and left in the tissue for 7 hours.} The blood and tissue were harvested immediately before the leaching study. The blood \textcolor{black}{and the tissue} samples with the fibers were kept at 36$^{\circ}$C during the entire time of the leaching study and were frozen immediately after the fibers were removed. Inductively coupled plasma mass spectrometry (ICP-MS) was used (Brooks Applied Labs\textcolor{black}{; Bothell, WA, USA}) to analyze the platinum content of the samples. The results of all samples were within error identical to the reference sample of pure pig blood\textcolor{black}{/tissue}, demonstrating a lack of significant leaching.

\subsection{\textit{In vivo} results.}
\begin{figure}[ht]
\centering
\includegraphics[height=2in]{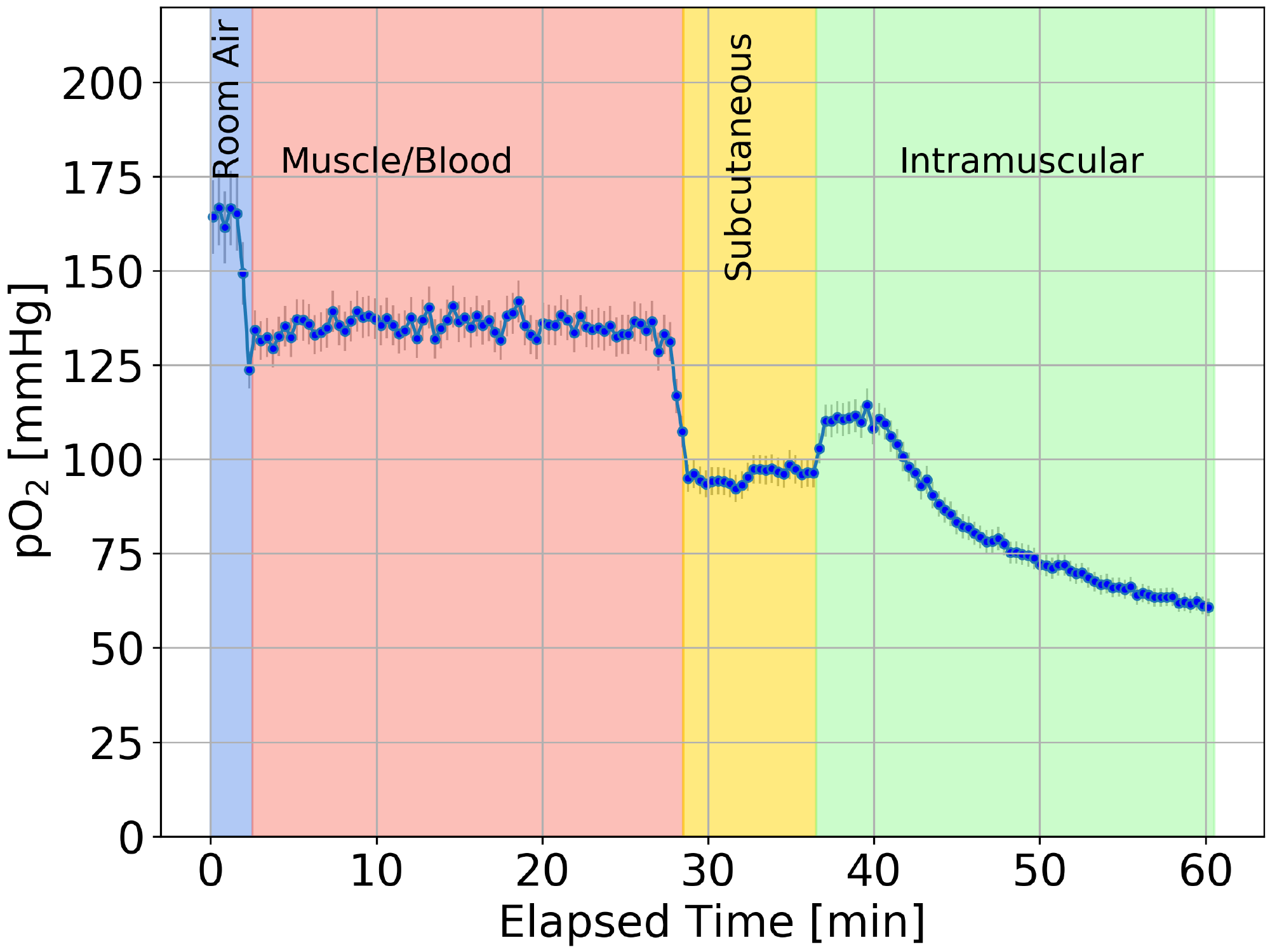}
\caption{Limb oxygenation following loss of cardiac function in a Yorkshire swine. The oxygen measurement was carried out in the hind limb (biceps femoris muscle) starting 1 minute after death. \textcolor{black}{The needle was subsequently inserted into muscle tissue, the subcutaneous tissue, and again into muscle tissue as shown in Figure~\ref{fig:PigMethods}.}}
\label{fig:PigData0}
\end{figure}

The results from the limb oxygenation measurement after loss of cardiac function are shown in Figure~\ref{fig:PigData0}.

The oxygen tension in a given tissue is a balance between both supply, via the vasculature, and demand, via the consumption of oxygen during cellular respiration. Muscle tissue is known to consume oxygen rapidly, as opposed to tissues such as the skin which very slowly utilize molecular oxygen. An overdose of Fatal Plus induces cardiac arrest, causing the cessation of pulsatile flow and therefore perfusion into tissue. It was expected that, when inserted into muscle, the probe would measure an exponential decay of oxygen tension, similar to that observed in prior studies \cite{Koolen2017-rt}.

The pO$_2$ measured at the first insertion location \textcolor{black}{ at the biceps femoris muscle} was found to be stable at 130~mmHg. \textcolor{black}{This value seems to be high at first, however can be explained by the high FiO$_2$ of 1. In addition, as the needle gauge of 18 was large and the needle was not smooth due to the drilled holes, the insertion of the needle may have induced bleeding at the sight of insertion resulting in the observed high pO$_2$ levels.\\ Following approximately 30 minutes of measurement, the needle was then moved and inserted into the skin above the biceps femoris muscle. Within the slowly oxygen-consuming skin, the prototype oxygen sensor measured a steady oxygen tension of 90~mmHg, a value in agreement with prior studies measuring subcutaneous oxygenation in pigs by \citeauthor{Ratnaraj2004-wb}. \cite{Ratnaraj2004-wb}\\
Following 8 minutes of measurement, the needle was re-inserted into the femoris muscle at a different location. At this second location, the oxygen partial pressure was observed to slowly decrease over the course of 20 minutes to a final value of approximately 60~mmHg.As this decrease in pO$_2$ was much slower than the equilibration time of the sensor, this decay was attributed to the slow metabolic rate of muscle tissue \cite{Ilkka2011-wk} and residual heart activity after Pentobarbital injection.\\
It is well known that intramuscular pO$_2$ is heterogeneous within a single muscle, varies between individuals\cite{Weick2016-zv}, and was shown to strongly depend on FiO$_2$ \cite{Paek2002-st,Liu2011-fy,Mahling2015-am}. In normal perfused tissue, \citeauthor{Doro2014-gu} measured values between 60~mmHg and 80~mmHg in canine models \cite{Doro2014-gu} at an FiO$_{2}$ of 0.5. \citeauthor{Weick2016-zv} measured an intramuscular pO$_2$ between 4.0~mmHg and 50.6~mmHg with an average difference of 19.9~mmHg at unknown FiO$_{2}$ \cite{Weick2016-zv}. \citeauthor{Paek2002-st} have measured a significant difference in intramuscular pO$_2$ in a rat model between FiO$_2$=0.21 (room air) and FiO$_2$=1, with pO$_2$=30-45~mmHg and 120-220~mmHg, respectively\cite{Paek2002-st}. Similar results were found by \citeauthor{Liu2011-fy} and \citeauthor{Mahling2015-am} in rats and mice, respectively \cite{Liu2011-fy,Mahling2015-am}. These values fall within agreement with our measurements in porcine muscle at FiO$_2$=1 immediately postmortem. It should be noted that conversations with large animal surgical staff confirmed these findings.}

\begin{figure}[ht]
\centering
\includegraphics[width=\linewidth]{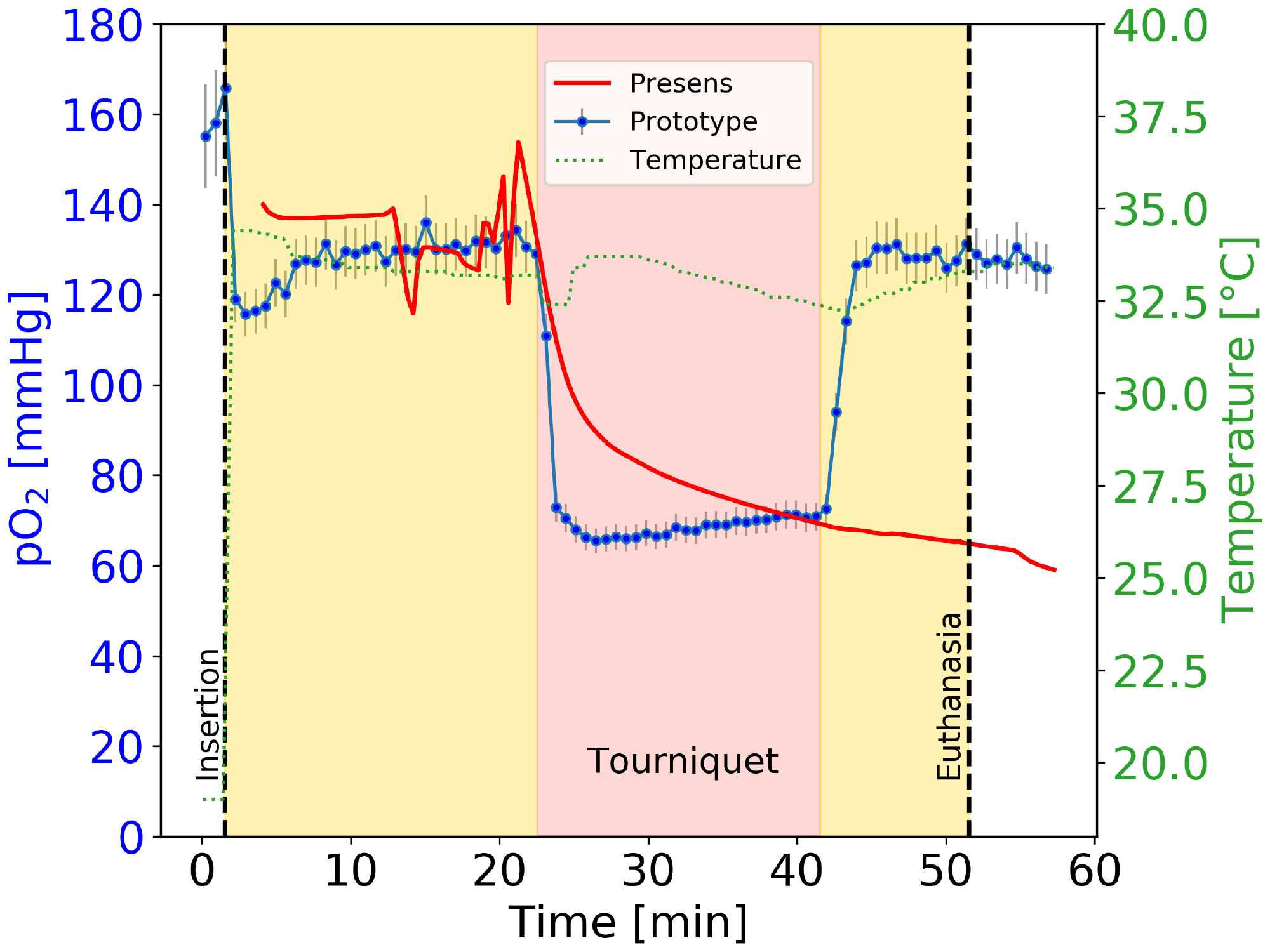}\\
\includegraphics[width=\linewidth]{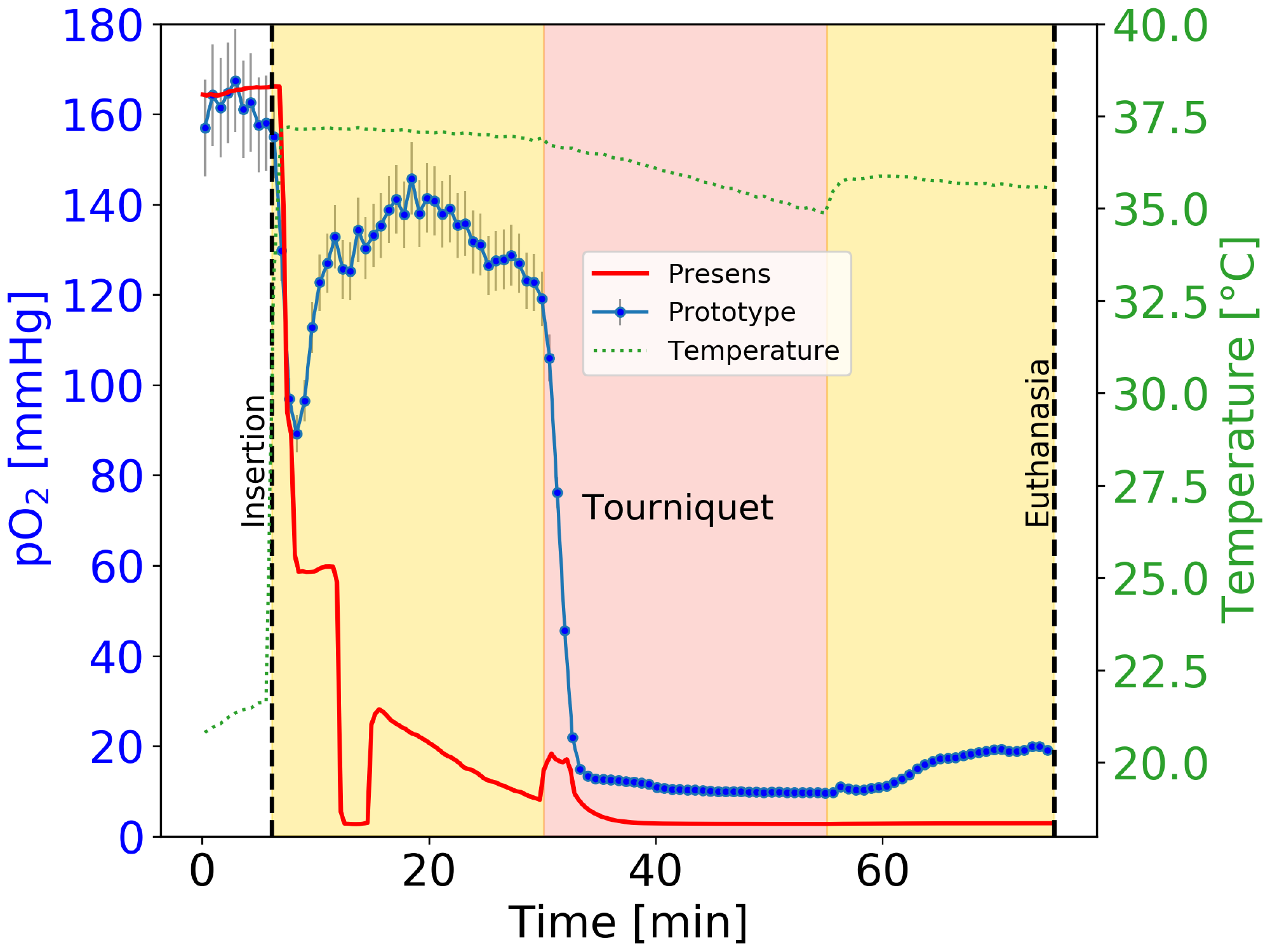}
\caption{\textcolor{black}{Limb oxygenation following the placement of a tourniquet. Top: pig 1 (Hampshire breed), needle-based measurement. Bottom: pig 2 (Yorkshire breed), catheter-based measurement.}}
\label{fig:PigData1}
\end{figure}

\textcolor{black}{The results from the \textit{in vivo} tourniquet model are shown in Figure~\ref{fig:PigData1}.} In both pigs, pO$_{2}$ values of 130~mmHg were measured before tourniquet application which is in agreement with the measurement in the first experiment \textcolor{black}{and with measurements in rat models at FiO$_2$=1 \cite{Paek2002-st,Liu2011-fy,Mahling2015-am}}. In both pigs, when the tourniquet was applied, the pO$_2$ dropped within less than a minute to 65~mmHg and 15~mmHg, respectively. In the first pig, the pO$_2$ immediately increased back to its initial value after tourniquet release whereas in the second pig the pO$_2$ only increased by approximately 10~mmHg to 20~mmHg over the course of 15 minutes.

The different pO$_{2}$ levels measured in the two pigs during and after tourniquet application can be explained by the different force with which the tourniquets were applied. The low pO$_2$ of only 20~mmHg in the second pig indicated severe tissue damage due to the applied force. \textcolor{black}{This is also in agreement with the observed change in skin color which was much more drastic in the second pig.}

In both pigs, the commercial oxygen sensor did not adequately reproduce the results measured with the prototype sensor. It is not clear why the commercial sensor shows this behaviour, however, it has to be noted that it was not designed for intramuscular nor any \textit{in vivo} applications and may have interacted with blood, was damaged, or rendered inert due to fouling. 

\section{CONCLUSIONS}
In order to measure deep tissue oxygenation, this study was focused on the development, construction, and validation of a fiber-based portable intramuscular oxygen sensing device. \textcolor{black}{TEOS sol-gel}, \textcolor{black}{3M Cavilon}, PEMA, and PPMA were evaluated as matrix materials to host a brightly emitting porphyrin oxygen sensor. \textcolor{black}{The Pt(II)-pivaloyl porphyrin in} PPMA was shown to have no humidity or pH dependence and exhibited a low photo-bleaching rate, which was essential for an \textcolor{black}{intramuscular} intensity-based \textcolor{black}{pO$_2$} measurement. 

The \textcolor{black}{PPMA} prototype sensors were tested in two different porcine models and showed an appropriate and reproducible response to changes in oxygenation. Subcutaneous measurements were in agreement with earlier measurements in rat models. Muscle oxygenation right after insertion of the sensor was measured to be around 130~mmHg, which can be explained by the high FiO$_2$ of 1 and additional bleeding due to needle insertion. In the future, bleeding could be reduced by using smaller gauge needles or by proper placing of the \textcolor{black}{needle/catheter} using ultrasound guidance. Both prototype sensors showed fast changes in pO$_2$ following tourniquet application. After release of the tourniquet, intramuscular pO$_2$ in the first pig increased rapidly to pre-tourniquet levels, whereas in the second pig the pO$_2$ only slowly increased by approximately 10~mmHg. This was most probably caused by tissue injury induced by an over-tightened tourniquet in the second pig. In the future, this could be controlled by applying a pneumatic tourniquet with 40-100~mmHg above the limb occlusion pressure \cite{AORN}, however, it is not straight-forward to fit a tourniquet to the leg shape of a pig while still having access to muscles. The commercial oxygen sensor used in parallel, which notably determines pO$_2$ based on the much more stable lifetime measurement, did not respond as expected \textit{in vivo}. 

Even-though in retrospective it would have been interesting to explore lower (or even several) values of FiO$_2$ it should be noted that the performance of the sensor is not largely affected by FiO$_2$. In addition, when using lower FiO$_2$ tissue pO$_2$ is expected to decrease to values where the sensor shows higher sensitivity and accuracy as shown in Figure~\ref{fig:2DCalib} and supporting Figure~S2.

While the needle-based version of the current prototype seems to be appropriate for single point measurements, due to its flexibility, the catheter-based version can be left in the tissue for a longer time. Thus, pO$_2$ values could be acquired over a long time frame so that an oxygen trend could be measured rather than an overall value. This could be advantageous since intramuscular oxygenation is expected to be heterogeneous inside a compartment. Furthermore, a single measurement could be enhanced by additional sensors arranged spatially in a grid, longitudinally along the probe, or both to acquire regional tissue pO$_2$ maps to understand the distribution of pO$_2$ for the improved diagnosis of ACS.\\ 
Whereas tissue oxygenation seems to be a much more physiologically relevant measurement for tissue health than total pressure, further investigations in a large animal compartment syndrome model are necessary to understand the possible benefit of the device for the assessment of compartment syndrome. In addition, a compartment syndrome model would give further insight into the effect of a change in total pressure on the oxygen partial pressure measurement. Measurements in a chamber where the total pressure was changed while keeping the ratio of nitrogen and air constant showed that changes in total pressure are directly related to changes in oxygen partial pressure (Supporting Figure~S8). However, it has to be noted that this model does not reflect the full complexity of the physiology during compartment syndrome and additional \emph{in vivo} studies will be needed in order to assess the interplay between total pressure and pO$_2$.

A new prototype where the intensity-measuring electronics is replaced with a lifetime-based setup is currently under construction. This will eliminate the need for bleaching corrections and will make the prototype less expensive and even more portable (matchbox-sized), which is necessary for an emergency setting where space and money are limited. \textcolor{black}{The lifetime-based measurement was seen to be more stable over time and thus exhibits potential to significantly increase the calibration stability of the sensor.}

For first-in human use the device needs further miniaturization in the needle gauge to decrease discomfort during insertion. The small 200~$\mu$m optical fiber used in the prototype is ideally suited for that and small gauge needles with sideports can be custom made by commercial producers. \textcolor{black}{To reduce the possibility of breaking, the empty space around the fiber in the needle could be filled up with epoxy resin so that the fiber tip is mounted flush with the epoxy resin. In addition, the sideports could be substituted by a porous housing material.} \\
Even though the fiber-part of the sensor will be a disposable device, for safe usage in humans a fabrication of the probe under sterile conditions or a sterilization of the oxygen sensor after fabrication will be necessary. As mentioned earlier, a sterilization process using EtO is favorable since it has been shown to be safe for fiber-based sensors and polymer coatings. The effects on the fiber and oxygen sensing coatings in our prototype still need to be investigated.
Cell toxicity assays will be used to further assess biocompatibilty of our oxygen sensor.

The device introduced in this paper was specifically developed to address the unmet need of early and adequate diagnosis of ACS. Due to the complex interplay of total pressure and pO$_2$ during ACS it is not clear whether the measurement of a single variable such as pO$_2$ is enough to secure the ACS diagnosis. Under specific circumstances, a scenario with high total pressure and low pO$_2$ could be indistinguishable from a scenario with low total pressure and high pO$_2$. In the future, we are planing to fully address this issue by adding a total pressure sensor to the same device. Also, this would allow physicians to get a direct comparison to the current diagnosis standard. To get a holistic understanding of this disease it might be even beneficial to measure additional parameters such as pH and lactate levels, in parallel in the same device. The measurement of these parameters in parallel could have additional utility in the monitoring and diagnosis of other hypoxia related diseases.

\begin{acknowledgement}

The authors thank Dr.~Joshua Tam, Dr.~Yakir Levin, and Prof.~Rox Anderson for their help in setting up the porcine studies. The authors also thank the Knight Surgery team for their assistance during the porcine studies.
This work was made possible by the U.S. Department of Defense through the Military Medical Photonics Program the AFOSR, FA9550-17-1-0277. The porcine studies were entirely funded off of internal funding.

\end{acknowledgement}

\begin{suppinfo}
\textcolor{black}{Data on linear temperature dependence; data on sensor accuracy; table summarizing different formulation properties; full oxygen calibration for all matrices; spectra from TEOS formulations; spectra on humidity sensitivity; spectra and lifetime from PPMA formulation; pH and total pressure dependence}.
\end{suppinfo}

\bibliography{biblio}

\providecommand{\latin}[1]{#1}
\makeatletter
\providecommand{\doi}
  {\begingroup\let\do\@makeother\dospecials
  \catcode`\{=1 \catcode`\}=2 \doi@aux}
\providecommand{\doi@aux}[1]{\endgroup\texttt{#1}}
\makeatother
\providecommand*\mcitethebibliography{\thebibliography}
\csname @ifundefined\endcsname{endmcitethebibliography}
  {\let\endmcitethebibliography\endthebibliography}{}
\begin{mcitethebibliography}{72}
\providecommand*\natexlab[1]{#1}
\providecommand*\mciteSetBstSublistMode[1]{}
\providecommand*\mciteSetBstMaxWidthForm[2]{}
\providecommand*\mciteBstWouldAddEndPuncttrue
  {\def\EndOfBibitem{\unskip.}}
\providecommand*\mciteBstWouldAddEndPunctfalse
  {\let\EndOfBibitem\relax}
\providecommand*\mciteSetBstMidEndSepPunct[3]{}
\providecommand*\mciteSetBstSublistLabelBeginEnd[3]{}
\providecommand*\EndOfBibitem{}
\mciteSetBstSublistMode{f}
\mciteSetBstMaxWidthForm{subitem}{(\alph{mcitesubitemcount})}
\mciteSetBstSublistLabelBeginEnd
  {\mcitemaxwidthsubitemform\space}
  {\relax}
  {\relax}

\bibitem[McQueen \latin{et~al.}(2000)McQueen, Gaston, and
  Court-Brown]{McQueen2000-ot}
McQueen,~M.~M.; Gaston,~P.; Court-Brown,~C.~M. Acute compartment syndrome.
  \emph{J. Bone Joint Surg. Br.} \textbf{2000}, \emph{82-B}, 200--203\relax
\mciteBstWouldAddEndPuncttrue
\mciteSetBstMidEndSepPunct{\mcitedefaultmidpunct}
{\mcitedefaultendpunct}{\mcitedefaultseppunct}\relax
\EndOfBibitem
\bibitem[Raza and Mahapatra(2015)Raza, and Mahapatra]{Raza2015-ey}
Raza,~H.; Mahapatra,~A. Acute compartment syndrome in orthopedics: causes,
  diagnosis, and management. \emph{Adv. Orthop.} \textbf{2015}, \emph{2015},
  543412\relax
\mciteBstWouldAddEndPuncttrue
\mciteSetBstMidEndSepPunct{\mcitedefaultmidpunct}
{\mcitedefaultendpunct}{\mcitedefaultseppunct}\relax
\EndOfBibitem
\bibitem[Gordon \latin{et~al.}(2018)Gordon, Talbot, Shero, Osier, Johnson,
  Balsamo, and Stockinger]{Gordon2018-ll}
Gordon,~W.~T.; Talbot,~M.; Shero,~J.~C.; Osier,~C.~J.; Johnson,~A.~E.;
  Balsamo,~L.~H.; Stockinger,~Z.~T. Acute Extremity Compartment Syndrome and
  the Role of Fasciotomy in Extremity War Wounds. \emph{Mil. Med.}
  \textbf{2018}, \emph{183}, 108--111\relax
\mciteBstWouldAddEndPuncttrue
\mciteSetBstMidEndSepPunct{\mcitedefaultmidpunct}
{\mcitedefaultendpunct}{\mcitedefaultseppunct}\relax
\EndOfBibitem
\bibitem[Tollens \latin{et~al.}(1998)Tollens, Janzing, and
  Broos]{Tollens1998-kb}
Tollens,~T.; Janzing,~H.; Broos,~P. The pathophysiology of the acute
  compartment syndrome. \emph{Acta Chir. Belg.} \textbf{1998}, \emph{98},
  171--175\relax
\mciteBstWouldAddEndPuncttrue
\mciteSetBstMidEndSepPunct{\mcitedefaultmidpunct}
{\mcitedefaultendpunct}{\mcitedefaultseppunct}\relax
\EndOfBibitem
\bibitem[von Keudell \latin{et~al.}(2015)von Keudell, Weaver, Appleton, Bae,
  Dyer, Heng, Jupiter, and Vrahas]{Von_Keudell2015-lq}
von Keudell,~A.~G.; Weaver,~M.~J.; Appleton,~P.~T.; Bae,~D.~S.; Dyer,~G. S.~M.;
  Heng,~M.; Jupiter,~J.~B.; Vrahas,~M.~S. Diagnosis and treatment of acute
  extremity compartment syndrome. \emph{Lancet} \textbf{2015}, \emph{386},
  1299--1310\relax
\mciteBstWouldAddEndPuncttrue
\mciteSetBstMidEndSepPunct{\mcitedefaultmidpunct}
{\mcitedefaultendpunct}{\mcitedefaultseppunct}\relax
\EndOfBibitem
\bibitem[Zhang \latin{et~al.}(2020)Zhang, Janssen, Tarabochia, von Keudell, and
  Chen]{Zhang2020-kn}
Zhang,~D.; Janssen,~S.~J.; Tarabochia,~M.; von Keudell,~A.; Chen,~N. Risk
  factors for death and amputation in acute leg compartment syndrome.
  \emph{Eur. J. Orthop. Surg. Traumatol.} \textbf{2020}, \emph{30},
  359--365\relax
\mciteBstWouldAddEndPuncttrue
\mciteSetBstMidEndSepPunct{\mcitedefaultmidpunct}
{\mcitedefaultendpunct}{\mcitedefaultseppunct}\relax
\EndOfBibitem
\bibitem[Gordon \latin{et~al.}(2018)Gordon, Talbot, Shero, Osier, Johnson,
  Balsamo, and Stockinger]{Gordon2018-cx}
Gordon,~W.~T.; Talbot,~M.; Shero,~J.~C.; Osier,~C.~J.; Johnson,~A.~E.;
  Balsamo,~L.~H.; Stockinger,~Z.~T. Acute Extremity Compartment Syndrome and
  the Role of Fasciotomy in Extremity War Wounds. \emph{Mil. Med.}
  \textbf{2018}, \emph{183}, 108--111\relax
\mciteBstWouldAddEndPuncttrue
\mciteSetBstMidEndSepPunct{\mcitedefaultmidpunct}
{\mcitedefaultendpunct}{\mcitedefaultseppunct}\relax
\EndOfBibitem
\bibitem[Mauffrey \latin{et~al.}(2019)Mauffrey, Hak, and
  Martin]{Mauffrey2019-sy}
Mauffrey,~C.; Hak,~D.~J.; Martin,~M.~P.,~III \emph{Compartment Syndrome: A
  Guide to Diagnosis and Management}; Springer Nature, 2019\relax
\mciteBstWouldAddEndPuncttrue
\mciteSetBstMidEndSepPunct{\mcitedefaultmidpunct}
{\mcitedefaultendpunct}{\mcitedefaultseppunct}\relax
\EndOfBibitem
\bibitem[Carrasco \latin{et~al.}(2015)Carrasco, Benito-Pe{\~n}a, Walt, and
  Moreno-Bondi]{Carrasco2015-vz}
Carrasco,~S.; Benito-Pe{\~n}a,~E.; Walt,~D.~R.; Moreno-Bondi,~M.~C. Fiber-optic
  array using molecularly imprinted microspheres for antibiotic analysis.
  \emph{Chem. Sci.} \textbf{2015}, \emph{6}, 3139--3147\relax
\mciteBstWouldAddEndPuncttrue
\mciteSetBstMidEndSepPunct{\mcitedefaultmidpunct}
{\mcitedefaultendpunct}{\mcitedefaultseppunct}\relax
\EndOfBibitem
\bibitem[Whitesides \latin{et~al.}(1975)Whitesides, Haney, Morimoto, and
  Harada]{Whitesides1975-jz}
Whitesides,~T.~E.; Haney,~T.~C.; Morimoto,~K.; Harada,~H. Tissue pressure
  measurements as a determinant for the need of fasciotomy. \emph{Clin. Orthop.
  Relat. Res.} \textbf{1975}, 43--51\relax
\mciteBstWouldAddEndPuncttrue
\mciteSetBstMidEndSepPunct{\mcitedefaultmidpunct}
{\mcitedefaultendpunct}{\mcitedefaultseppunct}\relax
\EndOfBibitem
\bibitem[Halanski \latin{et~al.}(2015)Halanski, Morris, Lee~Harper, and
  Doro]{Halanski2015-qt}
Halanski,~M.~A.; Morris,~M.~R.; Lee~Harper,~B.; Doro,~C. Intracompartmental
  Pressure Monitoring Using a Handheld Pressure Monitoring System. \emph{JBJS
  Essent Surg Tech} \textbf{2015}, \emph{5}, e6\relax
\mciteBstWouldAddEndPuncttrue
\mciteSetBstMidEndSepPunct{\mcitedefaultmidpunct}
{\mcitedefaultendpunct}{\mcitedefaultseppunct}\relax
\EndOfBibitem
\bibitem[McMillan \latin{et~al.}(2019)McMillan, Gardner, Schmidt, and
  Johnstone]{McMillan2019-bu}
McMillan,~T.~E.; Gardner,~W.~T.; Schmidt,~A.~H.; Johnstone,~A.~J. Diagnosing
  acute compartment syndrome-where have we got to? \emph{Int. Orthop.}
  \textbf{2019}, \emph{43}, 2429--2435\relax
\mciteBstWouldAddEndPuncttrue
\mciteSetBstMidEndSepPunct{\mcitedefaultmidpunct}
{\mcitedefaultendpunct}{\mcitedefaultseppunct}\relax
\EndOfBibitem
\bibitem[Guo \latin{et~al.}(2019)Guo, Yin, Jin, Zhang, Hou, and
  Zhang]{Guo2019-nc}
Guo,~J.; Yin,~Y.; Jin,~L.; Zhang,~R.; Hou,~Z.; Zhang,~Y. Acute compartment
  syndrome: Cause, diagnosis, and new viewpoint. \emph{Medicine} \textbf{2019},
  \emph{98}, e16260\relax
\mciteBstWouldAddEndPuncttrue
\mciteSetBstMidEndSepPunct{\mcitedefaultmidpunct}
{\mcitedefaultendpunct}{\mcitedefaultseppunct}\relax
\EndOfBibitem
\bibitem[Nelson(2013)]{Nelson2013-bo}
Nelson,~J.~A. Compartment pressure measurements have poor specificity for
  compartment syndrome in the traumatized limb. \emph{J. Emerg. Med.}
  \textbf{2013}, \emph{44}, 1039--1044\relax
\mciteBstWouldAddEndPuncttrue
\mciteSetBstMidEndSepPunct{\mcitedefaultmidpunct}
{\mcitedefaultendpunct}{\mcitedefaultseppunct}\relax
\EndOfBibitem
\bibitem[De~Santis and Singer(2015)De~Santis, and Singer]{De_Santis2015-fq}
De~Santis,~V.; Singer,~M. Tissue oxygen tension monitoring of organ perfusion:
  rationale, methodologies, and literature review. \emph{Br. J. Anaesth.}
  \textbf{2015}, \emph{115}, 357--365\relax
\mciteBstWouldAddEndPuncttrue
\mciteSetBstMidEndSepPunct{\mcitedefaultmidpunct}
{\mcitedefaultendpunct}{\mcitedefaultseppunct}\relax
\EndOfBibitem
\bibitem[Seekamp \latin{et~al.}(1998)Seekamp, Blankenburg, van Griensven, and
  Regel]{Seekamp1998-oi}
Seekamp,~A.; Blankenburg,~H.; van Griensven,~M.; Regel,~G. [Intramuscular pO2
  monitoring in compartment syndrome--an experimental study]. \emph{Zentralbl.
  Chir.} \textbf{1998}, \emph{123}, 285--91; discussion 291--2\relax
\mciteBstWouldAddEndPuncttrue
\mciteSetBstMidEndSepPunct{\mcitedefaultmidpunct}
{\mcitedefaultendpunct}{\mcitedefaultseppunct}\relax
\EndOfBibitem
\bibitem[Doro \latin{et~al.}(2014)Doro, Sitzman, and O'Toole]{Doro2014-gu}
Doro,~C.~J.; Sitzman,~T.~J.; O'Toole,~R.~V. Can intramuscular glucose levels
  diagnose compartment syndrome? \emph{J. Trauma Acute Care Surg.}
  \textbf{2014}, \emph{76}, 474--478\relax
\mciteBstWouldAddEndPuncttrue
\mciteSetBstMidEndSepPunct{\mcitedefaultmidpunct}
{\mcitedefaultendpunct}{\mcitedefaultseppunct}\relax
\EndOfBibitem
\bibitem[Weick \latin{et~al.}(2016)Weick, Kang, Lee, Kuether, Liu, Hansen,
  Kandemir, Rollins, and Mok]{Weick2016-zv}
Weick,~J.~W.; Kang,~H.; Lee,~L.; Kuether,~J.; Liu,~X.; Hansen,~E.~N.;
  Kandemir,~U.; Rollins,~M.~D.; Mok,~J.~M. Direct Measurement of Tissue
  Oxygenation as a Method of Diagnosis of Acute Compartment Syndrome. \emph{J.
  Orthop. Trauma} \textbf{2016}, \emph{30}, 585--591\relax
\mciteBstWouldAddEndPuncttrue
\mciteSetBstMidEndSepPunct{\mcitedefaultmidpunct}
{\mcitedefaultendpunct}{\mcitedefaultseppunct}\relax
\EndOfBibitem
\bibitem[Hansen \latin{et~al.}(2013)Hansen, Manzano, Kandemir, and
  Mok]{Hansen2013-yd}
Hansen,~E.~N.; Manzano,~G.; Kandemir,~U.; Mok,~J.~M. Comparison of tissue
  oxygenation and compartment pressure following tibia fracture. \emph{Injury}
  \textbf{2013}, \emph{44}, 1076--1080\relax
\mciteBstWouldAddEndPuncttrue
\mciteSetBstMidEndSepPunct{\mcitedefaultmidpunct}
{\mcitedefaultendpunct}{\mcitedefaultseppunct}\relax
\EndOfBibitem
\bibitem[Clark \latin{et~al.}(1953)Clark, Wolf, Granger, and
  Taylor]{Clark1953-se}
Clark,~L.~C.,~Jr; Wolf,~R.; Granger,~D.; Taylor,~Z. Continuous recording of
  blood oxygen tensions by polarography. \emph{J. Appl. Physiol.}
  \textbf{1953}, \emph{6}, 189--193\relax
\mciteBstWouldAddEndPuncttrue
\mciteSetBstMidEndSepPunct{\mcitedefaultmidpunct}
{\mcitedefaultendpunct}{\mcitedefaultseppunct}\relax
\EndOfBibitem
\bibitem[Vanderkooi \latin{et~al.}(1987)Vanderkooi, Maniara, Green, and
  Wilson]{Vanderkooi1987-at}
Vanderkooi,~J.~M.; Maniara,~G.; Green,~T.~J.; Wilson,~D.~F. An optical method
  for measurement of dioxygen concentration based upon quenching of
  phosphorescence. \emph{J. Biol. Chem.} \textbf{1987}, \emph{262},
  5476--5482\relax
\mciteBstWouldAddEndPuncttrue
\mciteSetBstMidEndSepPunct{\mcitedefaultmidpunct}
{\mcitedefaultendpunct}{\mcitedefaultseppunct}\relax
\EndOfBibitem
\bibitem[Rumsey \latin{et~al.}(1988)Rumsey, Vanderkooi, and
  Wilson]{Rumsey1988-jn}
Rumsey,~W.~L.; Vanderkooi,~J.~M.; Wilson,~D.~F. Imaging of phosphorescence: a
  novel method for measuring oxygen distribution in perfused tissue.
  \emph{Science} \textbf{1988}, \emph{241}, 1649--1651\relax
\mciteBstWouldAddEndPuncttrue
\mciteSetBstMidEndSepPunct{\mcitedefaultmidpunct}
{\mcitedefaultendpunct}{\mcitedefaultseppunct}\relax
\EndOfBibitem
\bibitem[Dunphy \latin{et~al.}(2002)Dunphy, Vinogradov, and
  Wilson]{Dunphy2002-fx}
Dunphy,~I.; Vinogradov,~S.~A.; Wilson,~D.~F. Oxyphor {R2} and G2: phosphors for
  measuring oxygen by oxygen-dependent quenching of phosphorescence.
  \emph{Anal. Biochem.} \textbf{2002}, \emph{310}, 191--198\relax
\mciteBstWouldAddEndPuncttrue
\mciteSetBstMidEndSepPunct{\mcitedefaultmidpunct}
{\mcitedefaultendpunct}{\mcitedefaultseppunct}\relax
\EndOfBibitem
\bibitem[Vinogradov \latin{et~al.}(1996)Vinogradov, Lo, Jenkins, Evans, Koch,
  and Wilson]{Vinogradov1996-ei}
Vinogradov,~S.~A.; Lo,~L.~W.; Jenkins,~W.~T.; Evans,~S.~M.; Koch,~C.;
  Wilson,~D.~F. Noninvasive imaging of the distribution in oxygen in tissue in
  vivo using near-infrared phosphors. \emph{Biophys. J.} \textbf{1996},
  \emph{70}, 1609--1617\relax
\mciteBstWouldAddEndPuncttrue
\mciteSetBstMidEndSepPunct{\mcitedefaultmidpunct}
{\mcitedefaultendpunct}{\mcitedefaultseppunct}\relax
\EndOfBibitem
\bibitem[Vinogradov \latin{et~al.}(1999)Vinogradov, Lo, and
  {others}]{Vinogradov1999-mv}
Vinogradov,~S.~A.; Lo,~L.~W.; {others}, Dendritic polyglutamic porphyrins:
  probing porphyrin protection by oxygen‐dependent quenching of
  phosphorescence. \emph{Chemistry--A European} \textbf{1999}, \emph{5},
  1338--1347\relax
\mciteBstWouldAddEndPuncttrue
\mciteSetBstMidEndSepPunct{\mcitedefaultmidpunct}
{\mcitedefaultendpunct}{\mcitedefaultseppunct}\relax
\EndOfBibitem
\bibitem[Stich \latin{et~al.}(2010)Stich, Fischer, and Wolfbeis]{Stich2010-om}
Stich,~M. I.~J.; Fischer,~L.~H.; Wolfbeis,~O.~S. Multiple fluorescent chemical
  sensing and imaging. \emph{Chem. Soc. Rev.} \textbf{2010}, \emph{39},
  3102--3114\relax
\mciteBstWouldAddEndPuncttrue
\mciteSetBstMidEndSepPunct{\mcitedefaultmidpunct}
{\mcitedefaultendpunct}{\mcitedefaultseppunct}\relax
\EndOfBibitem
\bibitem[Wang and Wolfbeis(2014)Wang, and Wolfbeis]{Wang2014-pi}
Wang,~X.-D.; Wolfbeis,~O.~S. Optical methods for sensing and imaging oxygen:
  materials, spectroscopies and applications. \emph{Chem. Soc. Rev.}
  \textbf{2014}, \emph{43}, 3666--3761\relax
\mciteBstWouldAddEndPuncttrue
\mciteSetBstMidEndSepPunct{\mcitedefaultmidpunct}
{\mcitedefaultendpunct}{\mcitedefaultseppunct}\relax
\EndOfBibitem
\bibitem[Roussakis \latin{et~al.}(2015)Roussakis, Li, Nichols, and
  Evans]{Roussakis2015-lv}
Roussakis,~E.; Li,~Z.; Nichols,~A.~J.; Evans,~C.~L. {Oxygen-Sensing} Methods in
  Biomedicine from the Macroscale to the Microscale. \emph{Angew. Chem. Int. Ed
  Engl.} \textbf{2015}, \emph{54}, 8340--8362\relax
\mciteBstWouldAddEndPuncttrue
\mciteSetBstMidEndSepPunct{\mcitedefaultmidpunct}
{\mcitedefaultendpunct}{\mcitedefaultseppunct}\relax
\EndOfBibitem
\bibitem[Roussakis \latin{et~al.}(2015)Roussakis, Li, Nowell, Nichols, and
  Evans]{Roussakis2015-vu}
Roussakis,~E.; Li,~Z.; Nowell,~N.~H.; Nichols,~A.~J.; Evans,~C.~L. Bright,
  ''Clickable'' Porphyrins for the Visualization of Oxygenation under Ambient
  Light. \emph{Angew. Chem. Int. Ed Engl.} \textbf{2015}, \emph{54},
  14728--14731\relax
\mciteBstWouldAddEndPuncttrue
\mciteSetBstMidEndSepPunct{\mcitedefaultmidpunct}
{\mcitedefaultendpunct}{\mcitedefaultseppunct}\relax
\EndOfBibitem
\bibitem[Stern and Volmer(1919)Stern, and Volmer]{Stern1919-mp}
Stern,~O.; Volmer,~M. {\"U}ber die Abklingungszeit der Fluoreszenz. \emph{Phys.
  Z.} \textbf{1919}, \emph{20}, 183--188\relax
\mciteBstWouldAddEndPuncttrue
\mciteSetBstMidEndSepPunct{\mcitedefaultmidpunct}
{\mcitedefaultendpunct}{\mcitedefaultseppunct}\relax
\EndOfBibitem
\bibitem[Koolen \latin{et~al.}(2017)Koolen, Li, Roussakis, Paul, Ibrahim,
  Matyal, Huang, Evans, and Lin]{Koolen2017-rt}
Koolen,~P. G.~L.; Li,~Z.; Roussakis,~E.; Paul,~M.~A.; Ibrahim,~A. M.~S.;
  Matyal,~R.; Huang,~T.; Evans,~C.~L.; Lin,~S.~J. {Oxygen-Sensing} {Paint-On}
  Bandage: Calibration of a Novel Approach in Tissue Perfusion Assessment.
  \emph{Plast. Reconstr. Surg.} \textbf{2017}, \emph{140}, 89--96\relax
\mciteBstWouldAddEndPuncttrue
\mciteSetBstMidEndSepPunct{\mcitedefaultmidpunct}
{\mcitedefaultendpunct}{\mcitedefaultseppunct}\relax
\EndOfBibitem
\bibitem[Li \latin{et~al.}(2018)Li, Marks, Evans, and Apiou-Sbirlea]{Li2018-mh}
Li,~Z.; Marks,~H.; Evans,~C.; Apiou-Sbirlea,~G. Sensing, monitoring, and
  release of therapeutics: the translational journey of next generation
  bandages. \emph{J. Biomed. Opt.} \textbf{2018}, \emph{24}, 1--9\relax
\mciteBstWouldAddEndPuncttrue
\mciteSetBstMidEndSepPunct{\mcitedefaultmidpunct}
{\mcitedefaultendpunct}{\mcitedefaultseppunct}\relax
\EndOfBibitem
\bibitem[Roussakis \latin{et~al.}(2020)Roussakis, Cascales, Marks, Li,
  Grinstaff, and Evans]{Roussakis2020-ks}
Roussakis,~E.; Cascales,~J.~P.; Marks,~H.~L.; Li,~X.; Grinstaff,~M.;
  Evans,~C.~L. Humidity-Insensitive Tissue Oxygen Tension Sensing for Wearable
  Devices†. \emph{Photochem. Photobiol.} \textbf{2020}, \emph{96},
  373--379\relax
\mciteBstWouldAddEndPuncttrue
\mciteSetBstMidEndSepPunct{\mcitedefaultmidpunct}
{\mcitedefaultendpunct}{\mcitedefaultseppunct}\relax
\EndOfBibitem
\bibitem[{Sayuri Ban, Ai Hosoki, Michiko Nishiyama, Atsushi Seki, Kazuhiro
  Watanabe}(2016)]{Sayuri-ls}
{Sayuri Ban, Ai Hosoki, Michiko Nishiyama, Atsushi Seki, Kazuhiro Watanabe},
  Optical fiber oxygen sensor using layer-by-layer stacked porous composite
  membranes. Proceedings Volume 9754, Photonic Instrumentation Engineering
  {III}. 2016\relax
\mciteBstWouldAddEndPuncttrue
\mciteSetBstMidEndSepPunct{\mcitedefaultmidpunct}
{\mcitedefaultendpunct}{\mcitedefaultseppunct}\relax
\EndOfBibitem
\bibitem[Chen \latin{et~al.}(2014)Chen, Formenti, McPeak, Obeid, Hahn, and
  Farmery]{Chen2014-oo}
Chen,~R.; Formenti,~F.; McPeak,~H.; Obeid,~A.~N.; Hahn,~C. E.~W.;
  Farmery,~A.~D. Optimizing Design for Polymer Fiber Optic Oxygen Sensors.
  \emph{IEEE Sens. J.} \textbf{2014}, \emph{14}, 3358--3364\relax
\mciteBstWouldAddEndPuncttrue
\mciteSetBstMidEndSepPunct{\mcitedefaultmidpunct}
{\mcitedefaultendpunct}{\mcitedefaultseppunct}\relax
\EndOfBibitem
\bibitem[Davenport \latin{et~al.}(2016)Davenport, Hickey, Phillips, and
  Kyriacou]{Davenport2016-fk}
Davenport,~J.~J.; Hickey,~M.; Phillips,~J.~P.; Kyriacou,~P.~A. Fiber-optic
  fluorescence-quenching oxygen partial pressure sensor using platinum
  octaethylporphyrin. \emph{Appl. Opt.} \textbf{2016}, \emph{55},
  5603--5609\relax
\mciteBstWouldAddEndPuncttrue
\mciteSetBstMidEndSepPunct{\mcitedefaultmidpunct}
{\mcitedefaultendpunct}{\mcitedefaultseppunct}\relax
\EndOfBibitem
\bibitem[Kocincova \latin{et~al.}(2007)Kocincova, Borisov, Krause, and
  Wolfbeis]{Kocincova2007-fs}
Kocincova,~A.~S.; Borisov,~S.~M.; Krause,~C.; Wolfbeis,~O.~S. Fiber-optic
  microsensors for simultaneous sensing of oxygen and pH, and of oxygen and
  temperature. \emph{Anal. Chem.} \textbf{2007}, \emph{79}, 8486--8493\relax
\mciteBstWouldAddEndPuncttrue
\mciteSetBstMidEndSepPunct{\mcitedefaultmidpunct}
{\mcitedefaultendpunct}{\mcitedefaultseppunct}\relax
\EndOfBibitem
\bibitem[Wolfbeis(2015)]{Wolfbeis2015-qp}
Wolfbeis,~O.~S. Luminescent sensing and imaging of oxygen: fierce competition
  to the Clark electrode. \emph{Bioessays} \textbf{2015}, \emph{37},
  921--928\relax
\mciteBstWouldAddEndPuncttrue
\mciteSetBstMidEndSepPunct{\mcitedefaultmidpunct}
{\mcitedefaultendpunct}{\mcitedefaultseppunct}\relax
\EndOfBibitem
\bibitem[Chen \latin{et~al.}(2013)Chen, Formenti, Obeid, Hahn, and
  Farmery]{Chen2013-vt}
Chen,~R.; Formenti,~F.; Obeid,~A.; Hahn,~C. E.~W.; Farmery,~A.~D. A fibre-optic
  oxygen sensor for monitoring human breathing. \emph{Physiol. Meas.}
  \textbf{2013}, \emph{34}, N71--81\relax
\mciteBstWouldAddEndPuncttrue
\mciteSetBstMidEndSepPunct{\mcitedefaultmidpunct}
{\mcitedefaultendpunct}{\mcitedefaultseppunct}\relax
\EndOfBibitem
\bibitem[Chen \latin{et~al.}(2016)Chen, Formenti, McPeak, Obeid, Hahn, and
  Farmery]{Chen2016-ri}
Chen,~R.; Formenti,~F.; McPeak,~H.; Obeid,~A.~N.; Hahn,~C.; Farmery,~A.
  Experimental investigation of the effect of polymer matrices on polymer fibre
  optic oxygen sensors and their time response characteristics using a vacuum
  testing chamber and a liquid flow apparatus. \emph{Sens. Actuators B Chem.}
  \textbf{2016}, \emph{222}, 531--535\relax
\mciteBstWouldAddEndPuncttrue
\mciteSetBstMidEndSepPunct{\mcitedefaultmidpunct}
{\mcitedefaultendpunct}{\mcitedefaultseppunct}\relax
\EndOfBibitem
\bibitem[Formenti \latin{et~al.}(2015)Formenti, Chen, McPeak, Murison,
  Matejovic, Hahn, and Farmery]{Formenti2015-od}
Formenti,~F.; Chen,~R.; McPeak,~H.; Murison,~P.~J.; Matejovic,~M.; Hahn,~C.
  E.~W.; Farmery,~A.~D. Intra-breath arterial oxygen oscillations detected by a
  fast oxygen sensor in an animal model of acute respiratory distress syndrome.
  \emph{Br. J. Anaesth.} \textbf{2015}, \emph{114}, 683--688\relax
\mciteBstWouldAddEndPuncttrue
\mciteSetBstMidEndSepPunct{\mcitedefaultmidpunct}
{\mcitedefaultendpunct}{\mcitedefaultseppunct}\relax
\EndOfBibitem
\bibitem[Chen \latin{et~al.}(2012)Chen, Farmery, Obeid, and Hahn]{Chen2012-qm}
Chen,~R.; Farmery,~A.~D.; Obeid,~A.; Hahn,~C. E.~W. A {Cylindrical-Core}
  {Fiber-Optic} Oxygen Sensor Based on Fluorescence Quenching of a Platinum
  Complex Immobilized in a Polymer Matrix. \emph{IEEE Sens. J.} \textbf{2012},
  \emph{12}, 71--75\relax
\mciteBstWouldAddEndPuncttrue
\mciteSetBstMidEndSepPunct{\mcitedefaultmidpunct}
{\mcitedefaultendpunct}{\mcitedefaultseppunct}\relax
\EndOfBibitem
\bibitem[Weiss \latin{et~al.}(1997)Weiss, Harrison, Feldman, and
  Brill]{Weiss1997-cr}
Weiss,~I.~K.; Harrison,~R.; Feldman,~J.~D.; Brill,~J.~E. Continuous arterial
  blood gas monitoring in pediatric patients: analysis of prolonged monitoring
  using the Paratrend 7 system. \emph{Crit. Care} \textbf{1997}, \emph{1},
  P114\relax
\mciteBstWouldAddEndPuncttrue
\mciteSetBstMidEndSepPunct{\mcitedefaultmidpunct}
{\mcitedefaultendpunct}{\mcitedefaultseppunct}\relax
\EndOfBibitem
\bibitem[Ye \latin{et~al.}(2007)Ye, Gao, Yin, and He]{Ye2007-lg}
Ye,~J.; Gao,~Z.; Yin,~J.; He,~Q. Hypoxia is a potential risk factor for chronic
  inflammation and adiponectin reduction in adipose tissue of ob/ob and dietary
  obese mice. \emph{Am. J. Physiol. Endocrinol. Metab.} \textbf{2007},
  \emph{293}, E1118--28\relax
\mciteBstWouldAddEndPuncttrue
\mciteSetBstMidEndSepPunct{\mcitedefaultmidpunct}
{\mcitedefaultendpunct}{\mcitedefaultseppunct}\relax
\EndOfBibitem
\bibitem[Yin \latin{et~al.}(2009)Yin, Gao, He, Zhou, Guo, and Ye]{Yin2009-rp}
Yin,~J.; Gao,~Z.; He,~Q.; Zhou,~D.; Guo,~Z.; Ye,~J. Role of hypoxia in
  obesity-induced disorders of glucose and lipid metabolism in adipose tissue.
  \emph{Am. J. Physiol. Endocrinol. Metab.} \textbf{2009}, \emph{296},
  E333--42\relax
\mciteBstWouldAddEndPuncttrue
\mciteSetBstMidEndSepPunct{\mcitedefaultmidpunct}
{\mcitedefaultendpunct}{\mcitedefaultseppunct}\relax
\EndOfBibitem
\bibitem[Al-Mutawa \latin{et~al.}(2018)Al-Mutawa, Herrmann, Corbishley, Losty,
  Phelan, and S{\'e}e]{Al-Mutawa2018-lr}
Al-Mutawa,~Y.~K.; Herrmann,~A.; Corbishley,~C.; Losty,~P.~D.; Phelan,~M.;
  S{\'e}e,~V. Effects of hypoxic preconditioning on neuroblastoma tumour
  oxygenation and metabolic signature in a chick embryo model. \emph{Biosci.
  Rep.} \textbf{2018}, \emph{38}\relax
\mciteBstWouldAddEndPuncttrue
\mciteSetBstMidEndSepPunct{\mcitedefaultmidpunct}
{\mcitedefaultendpunct}{\mcitedefaultseppunct}\relax
\EndOfBibitem
\bibitem[Crockett \latin{et~al.}(2019)Crockett, Cronin, Bommakanti, Chen, Hahn,
  Hedenstierna, Larsson, Farmery, and Formenti]{Crockett2019-iy}
Crockett,~D.~C.; Cronin,~J.~N.; Bommakanti,~N.; Chen,~R.; Hahn,~C. E.~W.;
  Hedenstierna,~G.; Larsson,~A.; Farmery,~A.~D.; Formenti,~F. Tidal changes in
  {PaO2} and their relationship to cyclical lung recruitment/derecruitment in a
  porcine lung injury model. \emph{Br. J. Anaesth.} \textbf{2019}, \emph{122},
  277--285\relax
\mciteBstWouldAddEndPuncttrue
\mciteSetBstMidEndSepPunct{\mcitedefaultmidpunct}
{\mcitedefaultendpunct}{\mcitedefaultseppunct}\relax
\EndOfBibitem
\bibitem[Jordan \latin{et~al.}(2004)Jordan, Kimpalou, Beghein, Dessy, Feron,
  and Gallez]{Jordan2004-kf}
Jordan,~B.~F.; Kimpalou,~J.~Z.; Beghein,~N.; Dessy,~C.; Feron,~O.; Gallez,~B.
  Contribution of oxygenation to {BOLD} contrast in exercising muscle.
  \emph{Magn. Reson. Med.} \textbf{2004}, \emph{52}, 391--396\relax
\mciteBstWouldAddEndPuncttrue
\mciteSetBstMidEndSepPunct{\mcitedefaultmidpunct}
{\mcitedefaultendpunct}{\mcitedefaultseppunct}\relax
\EndOfBibitem
\bibitem[Mahling \latin{et~al.}(2015)Mahling, Fuchs, Thaiss, Maier, Feger,
  Bukala, Harant, Eichner, Reutershan, Lang, Reischl, Pichler, and
  Kneilling]{Mahling2015-am}
Mahling,~M.; Fuchs,~K.; Thaiss,~W.~M.; Maier,~F.~C.; Feger,~M.; Bukala,~D.;
  Harant,~M.; Eichner,~M.; Reutershan,~J.; Lang,~F.; Reischl,~G.;
  Pichler,~B.~J.; Kneilling,~M. A Comparative pO2 Probe and
  [{18F]-Fluoro-Azomycinarabino-Furanoside} ({[18F]FAZA}) {PET} Study Reveals
  {Anesthesia-Induced} Impairment of Oxygenation and Perfusion in Tumor and
  Muscle. \emph{PLoS One} \textbf{2015}, \emph{10}, e0124665\relax
\mciteBstWouldAddEndPuncttrue
\mciteSetBstMidEndSepPunct{\mcitedefaultmidpunct}
{\mcitedefaultendpunct}{\mcitedefaultseppunct}\relax
\EndOfBibitem
\bibitem[Driessen \latin{et~al.}(2007)Driessen, Zarucco, Gunther, Burns, Lamb,
  Vincent, Boston, Jahr, and Cheung]{Driessen2007-kg}
Driessen,~B.; Zarucco,~L.; Gunther,~R.~A.; Burns,~P.~M.; Lamb,~S.~V.;
  Vincent,~S.~E.; Boston,~R.~A.; Jahr,~J.~S.; Cheung,~A. T.~W. Effects of
  low-volume hemoglobin glutamer-200 versus normal saline and arginine
  vasopressin resuscitation on systemic and skeletal muscle blood flow and
  oxygenation in a canine hemorrhagic shock model. \emph{Crit. Care Med.}
  \textbf{2007}, \emph{35}, 2101--2109\relax
\mciteBstWouldAddEndPuncttrue
\mciteSetBstMidEndSepPunct{\mcitedefaultmidpunct}
{\mcitedefaultendpunct}{\mcitedefaultseppunct}\relax
\EndOfBibitem
\bibitem[Carreau \latin{et~al.}(2011)Carreau, El~Hafny-Rahbi, Matejuk, Grillon,
  and Kieda]{Carreau2011-qv}
Carreau,~A.; El~Hafny-Rahbi,~B.; Matejuk,~A.; Grillon,~C.; Kieda,~C. Why is the
  partial oxygen pressure of human tissues a crucial parameter? Small molecules
  and hypoxia. \emph{J. Cell. Mol. Med.} \textbf{2011}, \emph{15},
  1239--1253\relax
\mciteBstWouldAddEndPuncttrue
\mciteSetBstMidEndSepPunct{\mcitedefaultmidpunct}
{\mcitedefaultendpunct}{\mcitedefaultseppunct}\relax
\EndOfBibitem
\bibitem[Tang \latin{et~al.}(2003)Tang, Tehan, Tao, and Bright]{Tang2003-qi}
Tang,~Y.; Tehan,~E.~C.; Tao,~Z.; Bright,~F.~V. {Sol-Gel-Derived} Sensor
  Materials That Yield Linear Calibration Plots, High Sensitivity, and
  Long-Term Stability. \emph{Anal. Chem.} \textbf{2003}, \emph{75},
  2407--2413\relax
\mciteBstWouldAddEndPuncttrue
\mciteSetBstMidEndSepPunct{\mcitedefaultmidpunct}
{\mcitedefaultendpunct}{\mcitedefaultseppunct}\relax
\EndOfBibitem
\bibitem[{Ismail} \latin{et~al.}(2013){Ismail}, {Mohd Zain}, and
  {Witjaksono}]{Ismail2013-wd}
{Ismail},~N.~H.; {Mohd Zain},~M.~N.; {Witjaksono},~G. Performance of
  TEOS/Octyl-triethoxylane/triton x-100 on
  tris(4,7'-diphenyl-1,10'-phenanthroline) Ruthenium (II) sol-gel for
  fluorescence-based optical sensor. 2013 IEEE 4th International Conference on
  Photonics (ICP). 2013; pp 278--280\relax
\mciteBstWouldAddEndPuncttrue
\mciteSetBstMidEndSepPunct{\mcitedefaultmidpunct}
{\mcitedefaultendpunct}{\mcitedefaultseppunct}\relax
\EndOfBibitem
\bibitem[Jorge \latin{et~al.}(2004)Jorge, Caldas, Rosa, Oliva, and
  Santos]{Jorge2004-ub}
Jorge,~P. A.~S.; Caldas,~P.; Rosa,~C.~C.; Oliva,~A.~G.; Santos,~J.~L. Optical
  fiber probes for fluorescence based oxygen sensing. \emph{Sens. Actuators B
  Chem.} \textbf{2004}, \emph{103}, 290--299\relax
\mciteBstWouldAddEndPuncttrue
\mciteSetBstMidEndSepPunct{\mcitedefaultmidpunct}
{\mcitedefaultendpunct}{\mcitedefaultseppunct}\relax
\EndOfBibitem
\bibitem[Yeh \latin{et~al.}(2006)Yeh, Chu, and Lo]{Yeh2006-hy}
Yeh,~T.-S.; Chu,~C.-S.; Lo,~Y.-L. Highly sensitive optical fiber oxygen sensor
  using {Pt(II}) complex embedded in sol--gel matrices. \emph{Sens. Actuators B
  Chem.} \textbf{2006}, \emph{119}, 701--707\relax
\mciteBstWouldAddEndPuncttrue
\mciteSetBstMidEndSepPunct{\mcitedefaultmidpunct}
{\mcitedefaultendpunct}{\mcitedefaultseppunct}\relax
\EndOfBibitem
\bibitem[Lo \latin{et~al.}(2008)Lo, Chu, Yur, and Chang]{Lo2008-rn}
Lo,~Y.-L.; Chu,~C.-S.; Yur,~J.-P.; Chang,~Y.-C. Temperature compensation of
  fluorescence intensity-based fiber-optic oxygen sensors using modified
  {Stern--Volmer} model. \emph{Sens. Actuators B Chem.} \textbf{2008},
  \emph{131}, 479--488\relax
\mciteBstWouldAddEndPuncttrue
\mciteSetBstMidEndSepPunct{\mcitedefaultmidpunct}
{\mcitedefaultendpunct}{\mcitedefaultseppunct}\relax
\EndOfBibitem
\bibitem[Chu and Lo(2007)Chu, and Lo]{Chu2007-ch}
Chu,~C.-S.; Lo,~Y.-L. High-performance fiber-optic oxygen sensors based on
  fluorinated xerogels doped with {Pt(II}) complexes. \emph{Sens. Actuators B
  Chem.} \textbf{2007}, \emph{124}, 376--382\relax
\mciteBstWouldAddEndPuncttrue
\mciteSetBstMidEndSepPunct{\mcitedefaultmidpunct}
{\mcitedefaultendpunct}{\mcitedefaultseppunct}\relax
\EndOfBibitem
\bibitem[Chu \latin{et~al.}(2011)Chu, Lo, and Sung]{Chu2011-go}
Chu,~C.-S.; Lo,~Y.-L.; Sung,~T.-W. Review on recent developments of fluorescent
  oxygen and carbon dioxide optical fiber sensors. \emph{Photonic Sensors}
  \textbf{2011}, \emph{1}, 234--250\relax
\mciteBstWouldAddEndPuncttrue
\mciteSetBstMidEndSepPunct{\mcitedefaultmidpunct}
{\mcitedefaultendpunct}{\mcitedefaultseppunct}\relax
\EndOfBibitem
\bibitem[Chu(2012)]{Chu2012-ho}
Chu,~C.~S. Optical Fiber Oxygen Sensor Based on Ru ({II}) Complex and Porous
  Silica Nanoparticles Embedded in {Sol-Gel} Matrix. \emph{Kemi} \textbf{2012},
  \emph{516}, 612--617\relax
\mciteBstWouldAddEndPuncttrue
\mciteSetBstMidEndSepPunct{\mcitedefaultmidpunct}
{\mcitedefaultendpunct}{\mcitedefaultseppunct}\relax
\EndOfBibitem
\bibitem[Zhang \latin{et~al.}(1991)Zhang, Bindra, Barrau, and
  Wilson]{Zhang1991-md}
Zhang,~Y.; Bindra,~D.~S.; Barrau,~M.~B.; Wilson,~G.~S. Application of cell
  culture toxicity tests to the development of implantable biosensors.
  \emph{Biosens. Bioelectron.} \textbf{1991}, \emph{6}, 653--661\relax
\mciteBstWouldAddEndPuncttrue
\mciteSetBstMidEndSepPunct{\mcitedefaultmidpunct}
{\mcitedefaultendpunct}{\mcitedefaultseppunct}\relax
\EndOfBibitem
\bibitem[Stolov \latin{et~al.}(2013)Stolov, Slyman, Burgess, Hokansson, Li, and
  Allen]{Stolov2013-fq}
Stolov,~A.~A.; Slyman,~B.~E.; Burgess,~D.~T.; Hokansson,~A.~S.; Li,~J.;
  Allen,~R.~S. Effects of sterilization methods on key properties of specialty
  optical fibers used in medical devices. Optical Fibers and Sensors for
  Medical Diagnostics and Treatment Applications {XIII}. 2013; p 857606\relax
\mciteBstWouldAddEndPuncttrue
\mciteSetBstMidEndSepPunct{\mcitedefaultmidpunct}
{\mcitedefaultendpunct}{\mcitedefaultseppunct}\relax
\EndOfBibitem
\bibitem[Hahn \latin{et~al.}(1993)Hahn, Alan, Bennett, and Hui]{Hahn1994-lf}
Hahn,~S.; Alan,~N.; Bennett,~M.; Hui,~H.~K. {CO2} Sensor using a hydrophilic
  polyurethane matrix. European Patent 0601816 A2, 1993\relax
\mciteBstWouldAddEndPuncttrue
\mciteSetBstMidEndSepPunct{\mcitedefaultmidpunct}
{\mcitedefaultendpunct}{\mcitedefaultseppunct}\relax
\EndOfBibitem
\bibitem[Gro(2016)]{Groupgetsgit}
Groupgets c12880ma arduino example.
  \url{https://github.com/groupgets/c12880ma/blob/master/arduino_c12880ma_example/arduino_c12880ma_example.ino},
  2016; Accessed: 2019-08-13\relax
\mciteBstWouldAddEndPuncttrue
\mciteSetBstMidEndSepPunct{\mcitedefaultmidpunct}
{\mcitedefaultendpunct}{\mcitedefaultseppunct}\relax
\EndOfBibitem
\bibitem[Lakowicz(2006)]{Lakowicz2006-ef}
Lakowicz,~J.~R. \emph{Principles of Fluorescence Spectroscopy}; Springer, 2006;
  Vol. Third Edition\relax
\mciteBstWouldAddEndPuncttrue
\mciteSetBstMidEndSepPunct{\mcitedefaultmidpunct}
{\mcitedefaultendpunct}{\mcitedefaultseppunct}\relax
\EndOfBibitem
\bibitem[Challa \latin{et~al.}(2017)Challa, Hargens, Uzosike, and
  Macias]{Challa2017-mg}
Challa,~S.~T.; Hargens,~A.~R.; Uzosike,~A.; Macias,~B.~R. Muscle Microvascular
  Blood Flow, Oxygenation, pH, and Perfusion Pressure Decrease in Simulated
  Acute Compartment Syndrome. \emph{J. Bone Joint Surg. Am.} \textbf{2017},
  \emph{99}, 1453--1459\relax
\mciteBstWouldAddEndPuncttrue
\mciteSetBstMidEndSepPunct{\mcitedefaultmidpunct}
{\mcitedefaultendpunct}{\mcitedefaultseppunct}\relax
\EndOfBibitem
\bibitem[Gershuni \latin{et~al.}(1985)Gershuni, Hargens, Lieber, O'Hara,
  Johansson, and Akeson]{Gershuni1985-dd}
Gershuni,~D.~H.; Hargens,~A.~R.; Lieber,~R.~L.; O'Hara,~R.~C.;
  Johansson,~C.~B.; Akeson,~W.~H. Decompression of an experimental compartment
  syndrome in dogs with hyaluronidase. \emph{Clin. Orthop. Relat. Res.}
  \textbf{1985}, 295--300\relax
\mciteBstWouldAddEndPuncttrue
\mciteSetBstMidEndSepPunct{\mcitedefaultmidpunct}
{\mcitedefaultendpunct}{\mcitedefaultseppunct}\relax
\EndOfBibitem
\bibitem[Ratnaraj \latin{et~al.}(2004)Ratnaraj, Kabon, Talcott, Sessler, and
  Kurz]{Ratnaraj2004-wb}
Ratnaraj,~J.; Kabon,~B.; Talcott,~M.~R.; Sessler,~D.~I.; Kurz,~A. Supplemental
  oxygen and carbon dioxide each increase subcutaneous and intestinal
  intramural oxygenation. \emph{Anesth. Analg.} \textbf{2004}, \emph{99},
  207--211\relax
\mciteBstWouldAddEndPuncttrue
\mciteSetBstMidEndSepPunct{\mcitedefaultmidpunct}
{\mcitedefaultendpunct}{\mcitedefaultseppunct}\relax
\EndOfBibitem
\bibitem[Ilkka \latin{et~al.}(2011)Ilkka, Bengt, Jukka, Sipil{\"a}, Vesa,
  Pirjo, Juhani, Kari, and Ylva]{Ilkka2011-wk}
Ilkka,~H.; Bengt,~S.; Jukka,~K.; Sipil{\"a},~H.~T.; Vesa,~O.; Pirjo,~N.;
  Juhani,~K.; Kari,~K.; Ylva,~H. Skeletal muscle blood flow and oxygen uptake
  at rest and during exercise in humans: a pet study with nitric oxide and
  cyclooxygenase inhibition. \emph{American Journal of Physiology-Heart and
  Circulatory Physiology} \textbf{2011}, \emph{300}, H1510--H1517\relax
\mciteBstWouldAddEndPuncttrue
\mciteSetBstMidEndSepPunct{\mcitedefaultmidpunct}
{\mcitedefaultendpunct}{\mcitedefaultseppunct}\relax
\EndOfBibitem
\bibitem[Paek \latin{et~al.}(2002)Paek, Chang, Brevetti, Rollins, Brady,
  Ursell, Hunt, Sarkar, and Messina]{Paek2002-st}
Paek,~R.; Chang,~D.~S.; Brevetti,~L.~S.; Rollins,~M.~D.; Brady,~S.;
  Ursell,~P.~C.; Hunt,~T.~K.; Sarkar,~R.; Messina,~L.~M. Correlation of a
  simple direct measurement of muscle pO(2) to a clinical ischemia index and
  histology in a rat model of chronic severe hindlimb ischemia. \emph{J. Vasc.
  Surg.} \textbf{2002}, \emph{36}, 172--179\relax
\mciteBstWouldAddEndPuncttrue
\mciteSetBstMidEndSepPunct{\mcitedefaultmidpunct}
{\mcitedefaultendpunct}{\mcitedefaultseppunct}\relax
\EndOfBibitem
\bibitem[Liu \latin{et~al.}(2011)Liu, Shah, Wilmes, Feiner, Kodibagkar,
  Wendland, Mason, Hylton, Hopf, and Rollins]{Liu2011-fy}
Liu,~S.; Shah,~S.~J.; Wilmes,~L.~J.; Feiner,~J.; Kodibagkar,~V.~D.;
  Wendland,~M.~F.; Mason,~R.~P.; Hylton,~N.; Hopf,~H.~W.; Rollins,~M.~D.
  Quantitative tissue oxygen measurement in multiple organs using {19F} {MRI}
  in a rat model. \emph{Magn. Reson. Med.} \textbf{2011}, \emph{66},
  1722--1730\relax
\mciteBstWouldAddEndPuncttrue
\mciteSetBstMidEndSepPunct{\mcitedefaultmidpunct}
{\mcitedefaultendpunct}{\mcitedefaultseppunct}\relax
\EndOfBibitem
\bibitem[{AORN Recommended Practices Committee}(2007)]{AORN}
{AORN Recommended Practices Committee}, Recommended practices for the use of
  the pneumatic tourniquet in the perioperative practice setting. \emph{AORN
  J.} \textbf{2007}, \emph{86}, 640--655\relax
\mciteBstWouldAddEndPuncttrue
\mciteSetBstMidEndSepPunct{\mcitedefaultmidpunct}
{\mcitedefaultendpunct}{\mcitedefaultseppunct}\relax
\EndOfBibitem
\end{mcitethebibliography}

\end{document}